\documentclass[aps,pra,twocolumn,superscriptaddress,floatfix,longbibliography]{revtex4-1}
\usepackage{amsfonts}
\usepackage{amssymb}
\usepackage{graphicx}
\usepackage{dcolumn}
\usepackage{bm}
\usepackage{amsmath}
\usepackage[colorlinks,linkcolor=magenta,citecolor=blue,urlcolor=blue]{hyperref}
\usepackage{changes}
\usepackage{float}

\newcommand{\ra}{\rangle}
\newcommand{\la}{\langle}

\begin{document}
\title{Statistical and dynamical aspects of quantum chaos in a kicked Bose-Hubbard dimer}
\author{Chenguang Liang}
\thanks{These authors contributed equally to this study.}
\affiliation{Beijing National Laboratory for Condensed Matter Physics, Institute
of Physics, Chinese Academy of Sciences, Beijing 100190, China}
\affiliation{School of Physical Sciences, University of Chinese Academy of Sciences,
Beijing, 100049, China}
\author{Yu Zhang}
\thanks{These authors contributed equally to this study.}
\affiliation{Beijing National Laboratory for Condensed Matter Physics, Institute
of Physics, Chinese Academy of Sciences, Beijing 100190, China}
\affiliation{School of Physical Sciences, University of Chinese Academy of Sciences,
Beijing, 100049, China}
\author{Shu Chen}
\email{schen@iphy.ac.cn}
\affiliation{Beijing National Laboratory for Condensed Matter Physics, Institute of
Physics, Chinese Academy of Sciences, Beijing 100190, China}
\affiliation{School of Physical Sciences, University of Chinese Academy of Sciences,
Beijing, 100049, China}
\date{\today }

\begin{abstract}
    Systems of interacting bosons in double-well potentials, modeled by two-site Bose-Hubbard models, are of significant theoretical and experimental interest and attracted intensive studies in contexts ranging from many-body physics and quantum dynamics to the onset of quantum chaos. In this work we systematically study a kicked two-site Bose-Hubbard model (Bose-Hubbard dimer) with the on-site potential difference being periodically modulated. Our model can be equivalently represented as a kicked Lipkin-Meshkov-Glick model and thus displays different dynamical behaviors from the kicked top model. By analyzing spectral statistics of Floquet operator, we unveil that the system undergoes a transition from regularity to chaos with increasing the interaction strength. Then based on semiclassical approximation and the analysis of R\'{e}nyi entropy of coherent states in the basis of Floquet operator eigenstates, we reveal the local chaotic features of our model, which indicate the existence of integrable islands even in the deep chaotic regime. The semiclassical analysis also suggests that the system in chaotic regime may display different dynamical behavior depending on the choice of initial states. Finally, we demonstrate that dynamical signatures of chaos can be manifested by studying dynamical evolution of local operators and out of time order correlation function as well as the entanglement entropy.  Our numerical results exhibit the richness of dynamics of the kicked Bose-Hubbard dimer in both regular and chaotic regimes as the initial states are chosen as coherent spin states located in different locations of phase space.
\end{abstract}

\maketitle

\section{Introduction}
 The two-mode or two-site Bose-Hubbard model is a prototype model to study many fundamental quantum dynamical phenomena, including the Landau-Zenner physics \cite{LZ1,LZ2,WuB}, Jospheson effect \cite{DW1,DW2}, self trapping \cite{Smerzi1999,Smerzi2001,FuLB}  and quantum chaos \cite{Strzys,Kidd,BHDimer2013,pencil1,pencil2}. A Bose gas trapped in a double-well potential or modulated optical lattice can be effectively described by a two-site Bose-Hubbard model (Bose-Hubbard dimer) \cite{Korsch,Chuchem,Leggett,Links,ZhouHQ}. With the experimental progress in cold atoms, the cold atomic system has become an ideal platform for exploring these intriguing dynamical phenomena \cite{Albiez,Levy,Exp-kicktop,Exp-kicktop2}.
As a minimum model for studying many-body dynamics, the Bose-Hubbard dimer has attracted intensive studies in the past years \cite{Korsch,Chuchem,Links,ZhouHQ}. Although exhibiting rich dynamical phenomena \cite{Tonel}, it is known that the Bose-Hubbard dimer is integrable \cite{Links,ZhouHQ} and thus does not support chaotic dynamics. Nevertheless, chaotic dynamical behavior can occur when the tunneling amplitude is periodically modulated \cite{BHDimer2013,Kidd}. If the periodical modulation of tunneling is applied in pulse, the model can be effectively described by a kicked top model, which is one of paradigmatic models widely used to understand quantum chaos \cite{Kicktop,Lombardi,Fox,Ghose,Kicktop1997,WangXG,Dorgra,Kidd2020,KickPtop,lerosepra,WangQ2021,WangQ2023}.

While the classical chaos is well defined as the  exponential sensitivity to the initial condition,  the canonical definition of quantum chaos is still on debate
due to the unitary property of quantum dynamics, especially for the quantum systems without classical correspondence \cite{Berry,WuB2021,Haake}.
A traditional approach to define and diagnosis quantum chaos, originated from the Bohigas-Giannoni-Schmit conjecture \cite{BGS}, is level spacing statistics. It allows us to identify a given quantum system as chaotic system when its level spacing statistics closely match the theoretical predictions derived from random matrix theory (RMT)\cite{Berry-1977spectrum,BGS,PhysRep,Huse,Atas,Guhr}. Besides the spectral statistics, the statistics of eigenvectors also provide useful signatures for detecting quantum chaos \cite{Berry1977,Zyczkowski,PhysRep,Haake1990,Kus,WangQ2023,WangQ2021}. 
In last decades, inspired by the huge progress in the quantum experimental techniques and the theory of quantum information,
the quantum chaos attracted much renewed attention from many different views, including the development of new tools for detecting quantum chaos and the application of quantum chaos to the forefront of research. Important diagnostic tools include the entanglement \cite{WangXG,Lombardi,Ghose,znidaric08,Huse13,bianchijhep,bianchipra,Nahum,Fazio} and the out-of-time-ordered correlator (OTOC) \cite{Larkin,Kitaev,Maldacena}, which provide alternative ways to pursue signatures of chaos from the dynamics of quantum information.Recent studies on various systems of interest have proven the OTOC to be a useful tool for studying quantum chaos and thermalization in quantum systems \cite{Larkin,Kitaev,Maldacena,Swingle,Swingle-pra,Xu,OTOC2017,OTOC2022,OTOC19,JPD,Trunin}. Although the exponential growth of OTOCs, or fast scrambling of quantum correlations, on short timescale is commonly considered as an indicator of chaos in many quantum systems \cite{OTOC2017,OTOC2022,OTOC19}, recent studies unveil the existence some counterexamples, which exhibit exponential growth of OTOCs resulting from unstable fixed points \cite{Richter,Fazio,CaoX,Galitski2020,Richter2023,Cameo,Kidd2021}. While the OTOCs in the periodically modulated Bose-Hubbard dimer are found to display obviously distinct behaviors in the chaotic and regular regimes \cite{Kidd2020},  scrambling dynamical behaviors around the saddle point of Bose-Hubbard dimers are also studied in Ref.\cite{Richter2023,Kidd2021}.

In this work, we study a generalized model of the kicked Bose-Hubbard dimer with the on-site potential difference being modulated periodically and explore the signatures of chaos by analyzing the spectrum and eigenvector statistics of the Floquet operator and the dynamical behaviors of OTOC and entanglement entropy. While the spectral statistics of Floquet operator indicates the existence of transition from regularity to chaos with increasing the interaction strength,  the analysis of R\'{e}nyi entropy of coherent states in the basis of Floquet operator eigenstates reveals distribution features in phase space,  resembling the semiclassical Poincar\'{e} section. Insights from the semiclassical correspondence imply that our system
displays fruitful dynamical phenomena in various parameter regimes with different initial states, which are witnessed by observing dynamical behaviors of local operators, OTOC and entanglement entropy.
We note that our model can not be reduced to the kicked top model, which can be realized by modulating the tunneling or the interaction strength of the kicked Bose-Hubbard dimer, and thus displays different dynamics from the previous studied models \cite{Kicktop,Lombardi,Fox,Ghose,Kicktop1997,WangXG,Dorgra,Kidd2020,KickPtop,lerosepra,WangQ2021,WangQ2023}.

The remainder of the paper is structured as follows.
In Sec.~\ref{sec:models}, we introduce the model. In sec.~\ref{sec:spectral}, we use the spectral statistics of Floquet unitary operator to show transitions from regular dynamics to quantum chaos at different parameters. In sec.~\ref{sec:phasespace}, we first present the Husimi distribution function of Floquet eigenstates. Based on the semiclassical approximation, we give the Poincar\'{e} section of systems. Then we study the phase space localization measure by means of multifractality dimensions of coherent state. Based on knowledge in previous section, we study the quantum dynamics with typical initial states in sec.~\ref{sec:quantumdynmics}. In sec.~\ref{sec:outlook}, we give a summary and outlook.

\section{Model}
\label{sec:models}

We consider the kicked Bose-Hubbard dimer model described  by
\begin{eqnarray}
    H&=& \nu (b_1^\dagger b_2+b_2^\dagger b_1)+\frac{U}{N}\sum_{i=1}^2n_i(n_i-1) \nonumber \\
    && +\frac{\mu}{2}(n_1-n_2)\sum_t \delta(t-n\tau), \label{BHD}
\end{eqnarray}
where $n_i= b^\dagger_i b_i$, $b_i$ and $b^\dagger_j$ are the boson destruction and creation operators fulfilling the commutation
\begin{equation}
     [b_i,b^\dagger_j]=\delta_{ij},[b^\dagger_i,b^\dagger_j]=[b_i,b_j]=0,
 \end{equation}
$\nu$ is the hopping amplitude between two sites, $U$ is the on site interaction strength, and $\mu$ denotes the periodically applied potential difference between two wells with the period $\tau$.

Using the boson operator, we can define the angular momentum operators \begin{align}
      & J_x=\frac{b_1^\dagger b_2+b_2^\dagger b_1}{2}   \\
      &J_y=\frac{i(b_1^\dagger b_2-b_2^\dagger b_1)}{2} \\
      &J_z=\frac{b_1^\dagger b_1-b_2^\dagger b_2}{2}.
  \end{align}
 where $J_a(a=x,y,z)$ are the components of angular momentum operator $\hat{J}$. One can check they satisfy the commutation relation$[J_i,J_j]=i\epsilon_{ijk} J_k$ and $\hat{J}^2=\frac{N}{2}\left(\frac{N}{2}+1\right)$. Since the total particle number $\hat{N}=n_1+n_2$ is a conserved quantity, after omitting some constants, the model (\ref{BHD}) can be represented as
 \begin{equation}
     H=2J_x+\frac{k}{2J}J_z^2+\mu J_z\sum_t \delta(t-n\tau), \label{hami1}
  \end{equation}
 where we have fixed $\hbar=1$ and set $\nu=1$ as the unit of energy, $J=\frac{N}{2}$ and $k=2U$. Our model is different from the celebrated kicked-top model \cite{Kicktop,Lombardi,Fox,Ghose,Kicktop1997,WangXG}, which reads as
 \begin{equation}
 H=\frac{k}{2J}J_z^2+\mu J_x\sum_t \delta(t-n\tau).
 \end{equation}
In comparison with kicked-top model, our model does not have the parity symmetry $P=e^{i\pi (J_x+J)}$ except when $\mu=0$ and $\pi$. Instead, it more resembles the kicked Lipkin-Meshkov-Glick (LMG) model \cite{Fazio,LMG2017} if we focus on the totally symmetric space.

\section{Spectrum statistics of Floquet operator}
\label{sec:spectral}
The time evolution operator corresponding the Hamiltonian (\ref{hami1}) is the unitary Floquet operator :\begin{equation}
U=e^{-i\mu J_z}e^{-i (2J_x+\frac{k}{2J} J_z^2) \tau}  \label{Uoperator}
\end{equation}
and we choose $\tau=1$ for convenience. The spectral statistics for a periodically driven quantum system can be carried out by analyzing the quasienergies (or eigenphases) of the Floquet operator. Next we study the spectral statistics of the unitary Floquet operator to show the transition from regular to chaos. The eigenphases of the Floquet operator $U$ are defined as
\[
U |\Phi_i \rangle= e^{i \omega_i} |\Phi_i \rangle,
\]
where $ \omega_i$ denotes the ith eigenphase of $U$ with corresponding eigenstate  $|\Phi_i \rangle$. As $\{\omega_i\}$ are $2 \pi$ periodic, the values of $\omega_i$ are restricted within the principal range $[-\pi, \pi)$.

\begin{figure}[h]
    \centering
    \includegraphics[width=0.5\textwidth,height=0.4\textwidth]{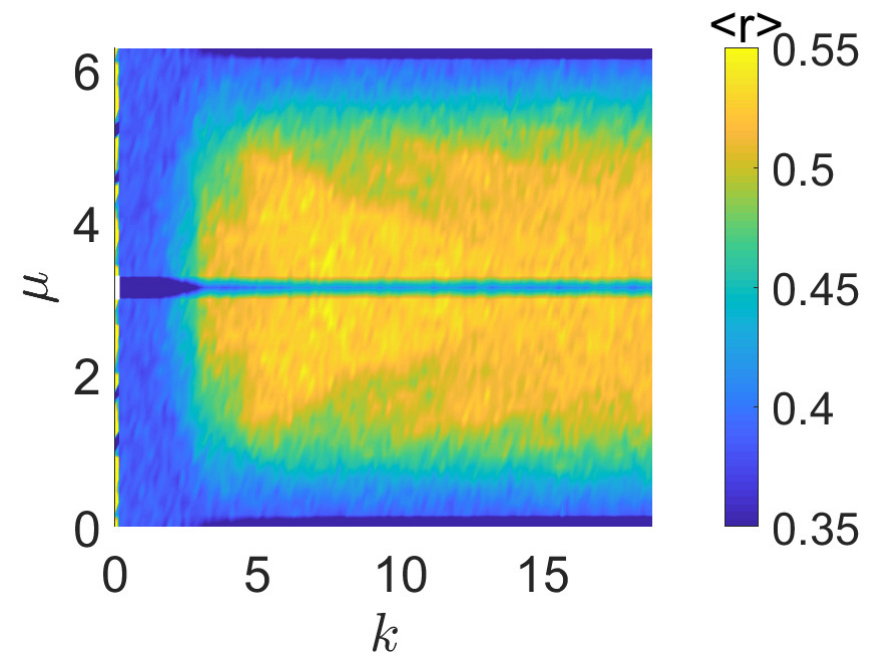}
    \caption{Mean level spacing ratio $\langle r \rangle$ as a function of $k$ and $\mu$ . When $k$ is small, the system is always in regular regime for any $u$. With the increase of $\mu$, periodical regular-chaotic crossover appears in regime with larger $k$. Here we take $J=1000$.}
    \label{ls1}
\end{figure}

\begin{figure}[h]
    \centering
    \includegraphics[width=0.52\textwidth,height=0.4\textwidth]{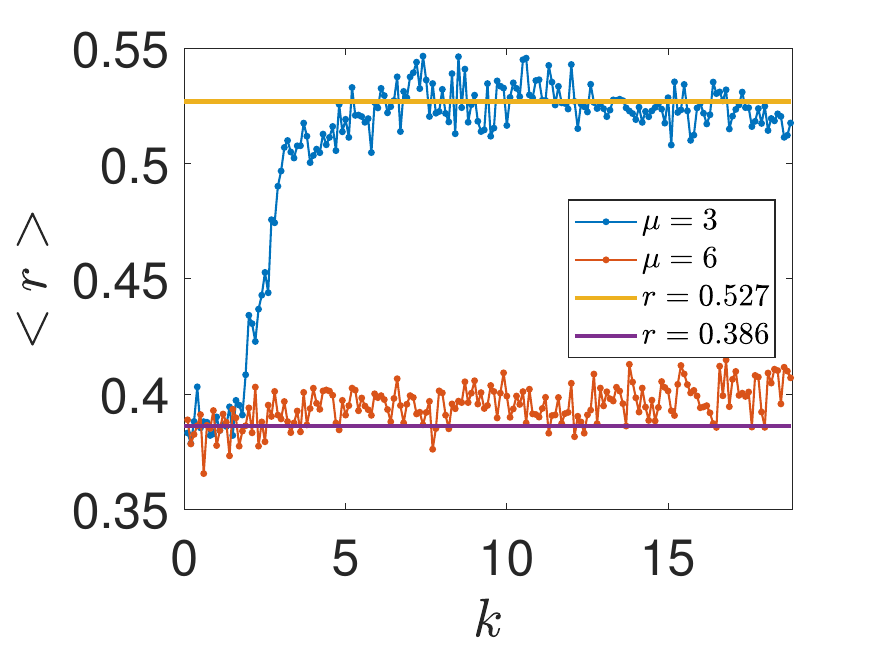}
    \caption{Mean level spacing ratio  $\langle r \rangle$ as a function of $k$ at fixed $\mu=3$ and $\mu=6$, respectively. }
    \label{ls2}
\end{figure}

The level spacing ratios $r_n$ are defined as
$$
r_n=\frac{\min(d_n,d_{n+1})}{\max(d_n,d_{n+1})}
$$
where $d_n=\omega_{n+1}-\omega_{n}$ is the spacing between two successive levels with $\omega_{n}$ being the nth eigenphase of the Floquet operator. The crossover from regular (integrable) to chaos in the model can be diagnosed by the mean level spacing ratio (MLSR) $\la r \ra$, defined as
\begin{equation}
    \langle r \rangle=\frac{1}{\mathcal{N}-1}\sum_{n=1}^{\mathcal{N}-1}r_n,
\end{equation}
where $\mathcal{N}$ is  the dimensions of Hilbert space of $U$. For regular systems, the Poisson statistics yields $\langle r \rangle_{P}\approx 0.386$, while for chaotic systems which fulfill Wigner-Dyson statistics, the mean value $\langle r \rangle_{WD}\approx 0.53$ \cite{Huse,Atas}. Since $\langle r \rangle$ is periodic as $\mu$ changes $2\pi$, in Fig.\ref{ls1}, we display $\langle r \rangle$ in the parameter space spanned by $k$ and \textcolor{red}{$\mu\in(0,2\pi]$}.
When $\mu$ is close to $2\pi$, the value of $\langle r \rangle$ indicates it belonging to the integrable regime. Varying $\mu$  and increasing $k$ can induce the crossover from integrable to chaotic regime. To see it clearly, in Fig.\ref{ls2} we show how $\langle r \rangle$ changes at different $k$ with fixed $\mu=3$ and $6$. It is clear that the model is always in the integrable regime when $\mu=6$. In the following discussion, we will focus on the case $\mu=3$ with different $k$.

\section{Husimi distributions of eigenstates and phase space localization measure}
\label{sec:phasespace}

Even though the level statistics of dynamical systems becomes a standard probe in the studies of quantum chaos, it cannot detect the local chaotic features in quantum systems.
In order to get a straightforward understanding of the onset of chaos, it is instructive to establish links between the properties of eigenstates of Floquet operator and the trajectories of
classically chaotic systems.  To this end, we shall first display Husimi distributions for various eigenstates of Floquet operator, which reflect the quantum phase-space distributions of Floquet  eigenstates, in the regular and chaotic regimes, respectively.  Then we show the phase space Poincar\'{e} section under the semiclassical approximation and reveal local chaotic features via the study of the phase space localization measure by  means of multifractality dimensions of coherent state.
\begin{figure*}
  \centering
  \includegraphics[width=1\textwidth,height=0.48\textwidth]{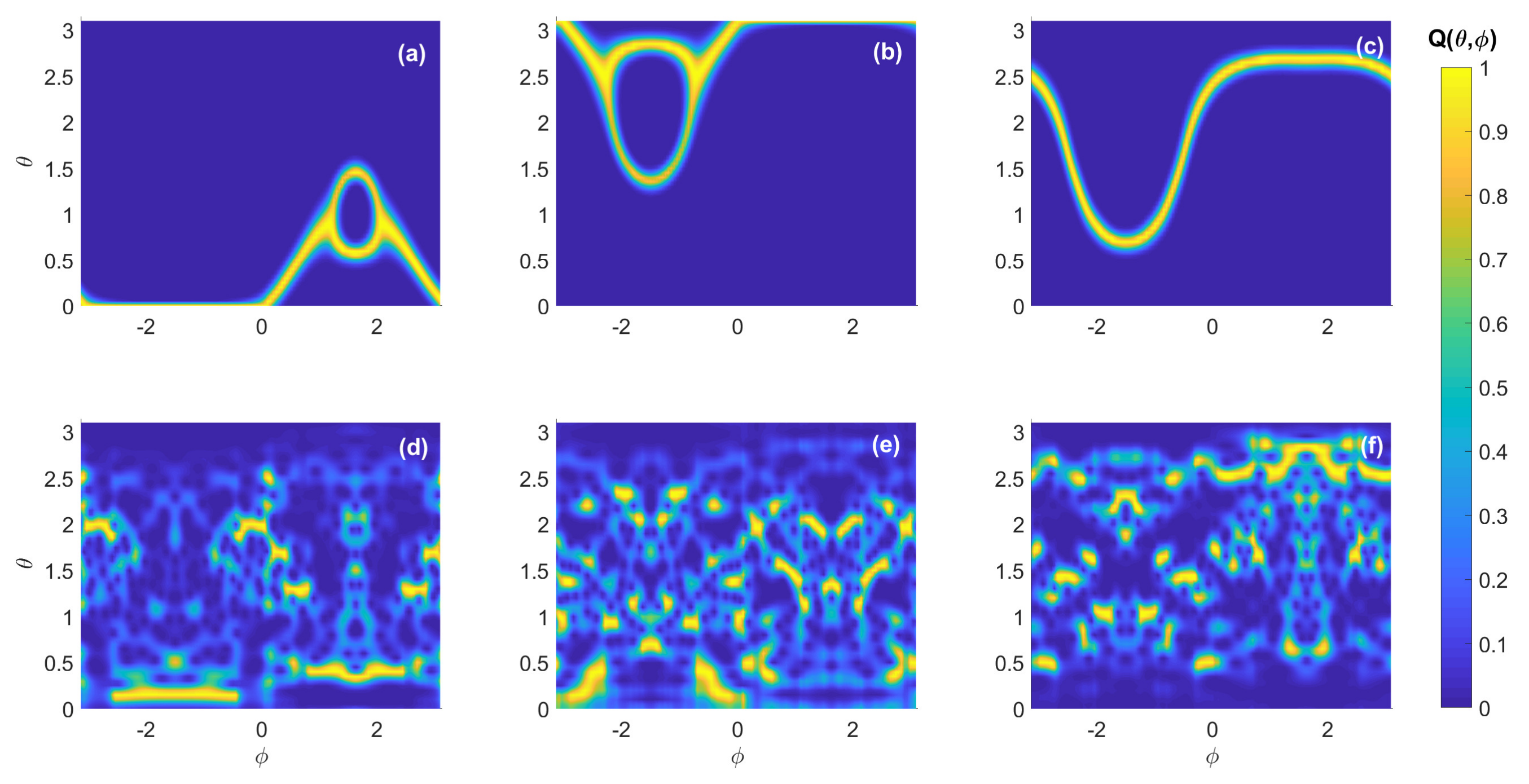}
  \caption{The first row: (a)-(c) are  Husimi function rescaled by its maximum value for three randomly chosen eigenstates of Floquet operator in the regular regime with parameters
  $\mu=3$ and $k=1$. The second row: (d)-(f) are  Husimi function rescaled by its maximum value for three randomly chosen eigenstates of Floquet operator in the chaotic regime with
  parameters $\mu=3$ and $k=8$. Here we fix $J=150$.}
  \label{husimi}
\end{figure*}

\subsection{Husimi distribution function}
As a kind of smoothed (coarsely grained) Wigner distribution \cite{Wigner,Husimi}, the Husimi distribution serves as a tool to reveal various aspects of eigenstates within the phase space \cite{ShiKJ,Takahashi}, offering insights into their localization properties. To define the Husimi function, we employ the generalized coherent spin states \cite{Coherentstate,WMZhang,SCS,SCS2} given by
\begin{align}
|\theta,\phi\rangle=\exp[i\theta(J_x\sin\phi-J_y\cos\phi)]|J,J\rangle \label{coherentstates},
\end{align}
where $\theta\in[0,\pi]$ and $\phi\in[-\pi,\pi)$, providing the orientation of $\hat{J}$. These coherent states satisfy
\[
\frac{1}{J}\langle \phi,\theta|\hat{J}|\phi,\theta\rangle=(\cos{\phi}\sin{\theta},\sin{\phi}\sin{\theta},\cos{\theta}).
\]
The Husimi function in the phase space for the $n$th eigenstate $|\Phi_n \rangle$ of the Floquet operator $U$ is given by
\begin{equation}
Q_n(\phi,\theta)=|\langle \theta,\phi|\Phi_n \rangle|^2.
\end{equation}

To gain an intuitive understanding, we depict the Husimi distribution of several randomly chosen eigenstates of Floquet operator for different values of $k$ in Fig.\ref{husimi}. In the regular regime with $k=1$ and $\mu=3$, the Husimi distributions display similar behaviors of regular periodic orbits, as shown in Figs.\ref{husimi}(a)-(c). It is clear that these eigenstates do not exhibit the parity symmetry in the phase space, as the parity corresponds to the operation of
$\theta \rightarrow \pi-\theta$ and $\phi \rightarrow 2\pi-\phi$.  On the other hand, in the chaotic regime, exemplified by the case with $k=8$ and $\mu=3$ as shown in Figs.\ref{husimi}(d)-(f)), the Husimi distribution in the phase space become random and irregular.
\begin{figure*}
    \centering
    \includegraphics[width=1\textwidth,height=0.56\textwidth]{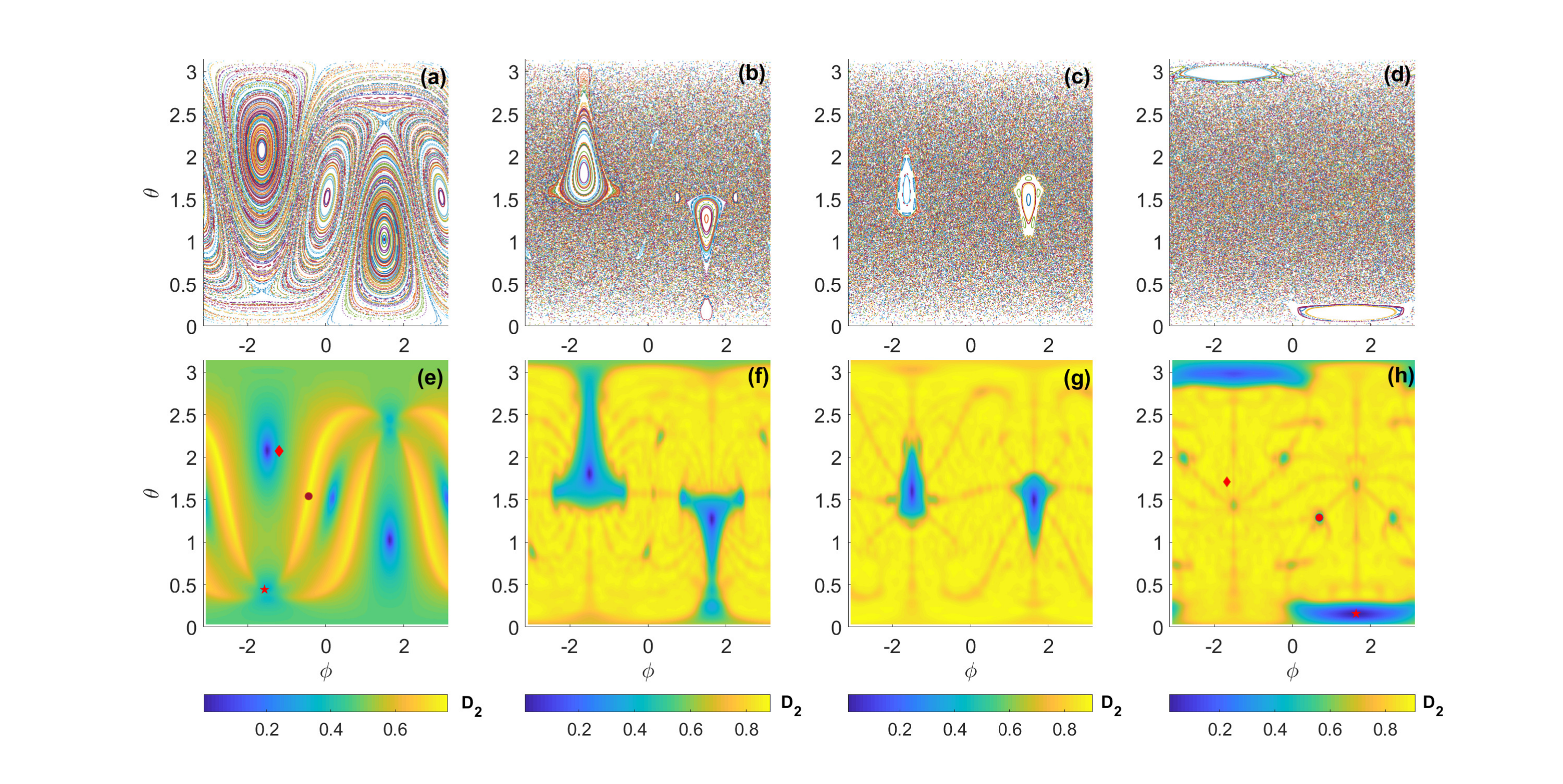}
    \caption{First row (a)-(d): phase space Poincar\'{e} section of the classic equation of motions. The initial states $\left(\theta,\phi\right)$ are selected on grid of 20 $\times$20 on
    the phase space. Each trajectory is evolved for 400 kicks. The parameter $\mu$ is fixed at 3, and $k=1,3,5$, and $8$ from (a) to (d), respectively. Second row (e)-(h): The multifractal
    dimension of coherent states, calculated on a grid of 100$\times$100 on the parameter space of $\theta$ and $\phi$. The parameter $\mu$ is fixed at 3, and  and $k=1,3,5$, and $8$ from
    (e) to (h), respectively. In (e) and (h), we marked the coherent states which we shall be utilized as initial states in the calculation of dynamics in sec.~\ref{sec:quantumdynmics}}
    \label{psren}
\end{figure*}

\subsection{Semiclassical approximation}
\label{sec:Poincare}
When $J \rightarrow \infty$, our model has a well defined quantum-classical correspondence, so the chaotic behaviour of the quantum system can be also manifested in its semiclassical dynamics.
Defining $\hat{m}=\hat{J}/J$, in the limit of $J\rightarrow \infty$, we see that  $\hat{m}$ behaves as classical angular momentum due to the vanishing of commutators between $m_\alpha$
($\alpha=x,y,z$). As the classical angular momentum $\hat{m}=(m_x,m_y,m_z)$ is a unit vector, it can be parameterized in terms of the azimuthal angle $\theta$ and polar angle $\phi$ as $\hat{m}=(\cos{\phi}\sin{\theta},\sin{\phi}\sin{\theta},\cos{\theta})$, where $\theta\in[0,\pi]$ and $\phi\in[-\pi,\pi)$. Hence, the classical phase space is a two-dimensional space with variables $\phi=\arctan{m_y/m_x}$ and $\theta=\arccos{m_z}$. We use the $\hat{m}$ to re-express Hamiltonian as
\begin{equation}
    H=J\left[ 2 m_x+\frac{k}{2}m_z^2+\mu m_z\sum_t \delta(t-n\tau)\right].
\end{equation}

The full dynamical map can be obtained by following two steps. Firstly we consider the evolution of time-independent part $H=J(2m_x+\frac{k}{2}m_z^2)$, the classical equation of motion of $\hat{m}$ is
\begin{align}
      & \dot{m_x}=-km_y m_z,    \\
      &\dot{m_y}=-2 m_z+km_x m_z,\\
      &\dot{m_z}=2 m_y,
\end{align}
where  the coefficient J can be canceled by the commutation of $\hat{m}$. Then the kicked part of the Floquet operator alone yields the map
\begin{align}
      &\Tilde{m}_x=\cos{\mu}m_x-\sin{\mu}m_y, \\
      &\Tilde{m}_y=\sin{\mu}m_x+\cos{\mu}m_y, \\
      &\Tilde{m}_z=m_z.
\end{align}

In the first row of Fig.(\ref{psren}), we present the Poincar\'{e} section for different values of the parameter $k$, while keeping $\mu$ fixed at $\mu=3$.These figures also indicate that the model does not have the exact parity symmetry any more. When $k=1$, as illustrated in Fig.\ref{psren}(a), the model is in the integrable regime, characterized by closed classical orbits and two approximate unstable fixed points. When the value of $k$ increases,  as exemplified by the cases of $k=3$ and $k=5$ shown in Figs. \ref{psren} (b) and \ref{psren}(c), respectively, two closed periodic orbits in the integrable regime vanish, and the remaining two gradually decrease in size. Finally, for the case of $k=8$ shown in Fig.\ref{psren} (d), corresponding to $\langle r \rangle \approx 0.53$, the system enters the quantum chaotic regime. In this regime, all periodic orbits in the bulk of the phase space disappear, except for a few very small isolated unstable fixed points (which shall be further  discussed in the next section). However, two integrable islands emerge in corner regions of phase space around $\theta \approx 0$ and $\pi$, respectively. We numerically check that these two integrable islands still survive when $k$ is very large. In the appendix ~\ref{sphere PS}, we also exhibit the Poincar\'{e} section on the sphere phase space using the same parameters utilized in Fig.\ref{psren}.

\subsection{Multifractality dimensions of coherent states}
\label{sec:coherntrenyi}
In order to further characterize the phase space structure and get more insight into the quantum-classical correspondence, we study the multifractal properties of the coherent states in this subsection. We still use the generalized coherent spin states given by Eq.\eqref{coherentstates}. To define the R\'{e}nyi entropy and multifractal dimensions of coherent states, we expand it in the orthonormal eigensates basis of the Floquet operators $U$
 \begin{equation}
     |\theta,\phi\rangle=\sum_{i=1}^{\mathcal{N}}  c_{i}|\Phi_i\rangle,
 \end{equation}
 where the $|\Phi_i\rangle$ is the $i$-th eigensates of Floquet operator and $c_{i}=\langle \Phi_i|\theta,\phi\rangle$. Then we can define the R\'{e}nyi entropy $S_q$ and multifractal dimensions $D_q$ \cite{Dq1,Dq2}:
\begin{equation}
   S_q=\frac{1}{1-q}\ln{\sum_{i=1}^{\mathcal{N}}|c_{i}|^{2q}} \quad  \text {and} \quad  D_q=\frac{S_q}{\ln\mathcal{N}} . \label{eq:renyi}
\end{equation}

For finite $\mathcal{N}$,  we have $D_q\in[0,1]$. The values of $D_q$
 decrease with increasing $q$ for $q\geq0$. The fractal dimensions of $\mathcal{D}_q^\infty$ are obtained via $\mathcal{D}_q^\infty=\lim_{\mathcal{N}\to\infty}D_q$ in the limit of $\mathcal{N}\to\infty$. The fractal dimensions measure the degree of ergodicity of a quantum state in Hilbert space. For a completely localized state $\mathcal{D}_q^\infty=0$ for $q>0$, whereas $\mathcal{D}_q^\infty=1$ denotes an extended state in the Hilbert space of Floquet operator and thus corresponds to an ergodic state.
 The multifractal states are the extended non-ergodic states and characterized by $0<\mathcal{D}_q^\infty<1$. In our calculation,  we choose $q=2$ with  $\mathcal{D}_2$ being the logarithm of the well-known participation ratio. In the second row of Fig.\ref{psren}, we depict $D_2$ as a function of $\phi$ and $\theta$ for different values of $k$. Comparing this with the first row of Fig.\ref{psren}, we observe a very similar structure to the Poincar\'{e} section discussed in Section~\ref{sec:Poincare}. Such a similarity suggests that the multifractal dimensions of coherent states exhibit an obvious correspondence to the underlying semiclassical Poincar\'{e} section and thus provide very useful information for understanding quantum chaos \cite{WangQ2021,Renyi-jpa,WangQ-Dicke}. While Figs.\ref{psren}(e) and (f) look like bearing a little resemblance to distributions with parity symmetry, we note that the resemblance is notably absent when $\mu$ deviates significantly from $\pi$.

 In the case of $k=1$ shown in Fig.\ref{psren} (e), the positions of $D_q\approx0$ are located at the regions corresponding to the most regular closed periodic orbits around the stable fixed points in the Poincar\'{e} sections, indicating the coherent states located at these points are the localized states in the Hilbert space, while the points with higher multifractal dimension are along the longer periodic orbits.
 As the $k$ increases, $D_2$ take larger values and exhibit an approximately uniform distribution in the rest of  phase space except for some integrable islands in the bulk, as shown in  Fig.\ref{psren} (f) and Fig.\ref{psren} (g). When $k$ enters the quantum chaotic regime as shown in Fig.\ref{psren} (h), where all integrable islands in the bulk disappear, the coherent states have high fractal dimensions almost everywhere in the bulk, but there are still some points with small values in the bulk, some of them will disappear when $J$ becomes large, while others are related to isolated fixed points in the semiclassical dynamics \cite{Kicktop,BHDimer2013}. These small patches are blurred in the background and need to be carefully found in the Poincar\'{e} section. A detailed analysis is given in the appendix~\ref{4p fixed point}. Finally, within large integrable islands near the south and north poles of the phase space, the fractal dimensions reach minimal values, even within chaotic regimes.

\begin{figure}[h]
    \centering
    \includegraphics[width=0.5\textwidth,height=0.4\textwidth]{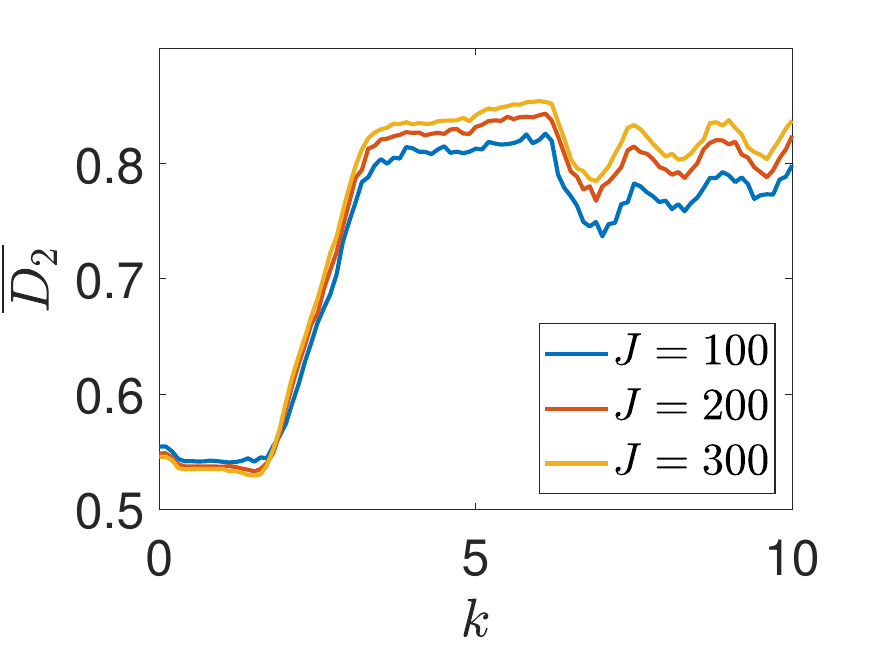}
    \caption{Average multifractal dimension $\bar{D_2}$ as a function of $k$ for different $J$. The parameter $\mu$ is fixed at $3$.}
    \label{amd}
\end{figure}

 We also define the average multifractal dimensions
 \[
 \bar{D_2}=\frac{1}{4\pi}\int dS D_2,
  \]
 which represent the average $D_2$ across the entire phase space. In Fig.\ref{amd}, we plot $\bar{D_2}$ versus $k$ for various $J=100$, $200$ and $300$. While $\bar{D_2}$ changes slowly with increasing $k$ for smaller $k$. Our results illustrate a rapid growth of $\bar{D_2}$ with increasing $k$ when $k>1.6$, resembling the behavior of the mean level spacing ratio presented in Fig.\ref{ls2}. However, as $k$ enters the quantum chaotic regime, with the gap ratio near to 0.53, $\bar{D_2}$ begins to exhibit oscillations. This distinct behavior from the mean level spacing ratio is attributed to the existence of integrable islands even for large $k$.

\begin{figure*}
  \centering
  \includegraphics[width=1\textwidth,height=0.48\textwidth]{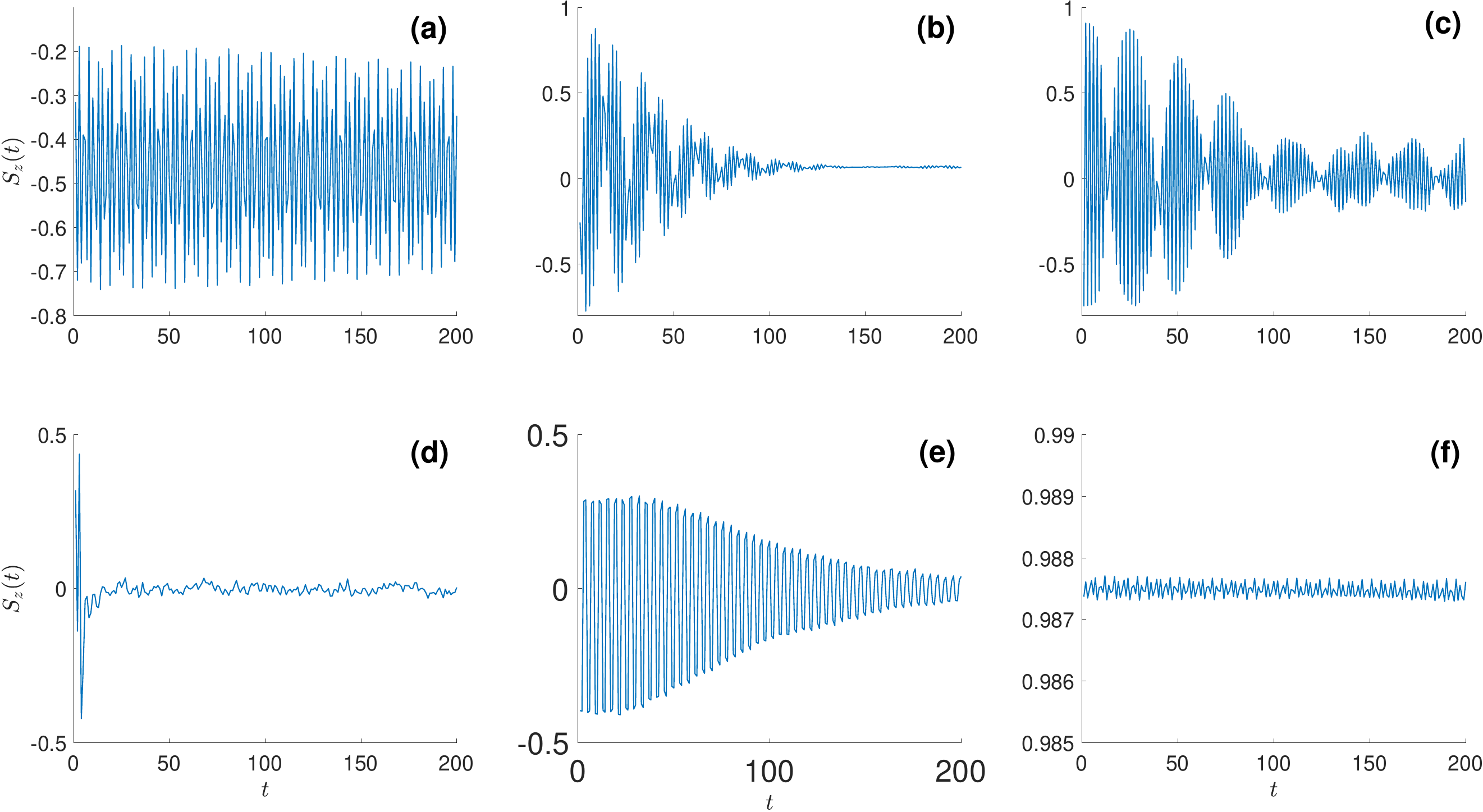}
  \caption{Dynamics of $S_z$ from different initial coherent states. The first row: (a)-(c) is in the regular regime with parameters $\mu=3$ and $k=1$. From left to right,  the initial states correspond to diamond, circle and star points marked in the Fig.\ref{psren} (e). The second row: (d)-(f) is in the chaotic regime with parameters $\mu=3$ and $k=8$. From left to right, the initial states correspond to diamond, circle and star points marked in the Fig.\ref{psren} (h). Here we fix $J=800$.}
  \label{Szt}
 \end{figure*}
\section{signatures from quantum dynamics}
\label{sec:quantumdynmics}
In the previous section, we have analyzed our system through its structure of spectral and eigenvector statistics as well as its semiclassical dynamics. We have conformed some interesting local dynamical features which can not  be captured by the statistics of spectrum of Floquet operator. Based on the analysis of phase space localization measure of eigenvector statistics and its corresponding classical phase space trajectories, we infer that the choice of different initial states may give rise to quite different quantum dynamics even for system with the same mean level spacing ratio of spectrum statistics. In this section we study the quantum dynamics to further display the feature of our model. For simplicity, we focus on the cases of $k=1$ and $k=8$.

\subsection{Dynamical evolution of local operators}
  Before the discussion for the quantum dynamical signature of chaos, we calculate the dynamical evolution of local operator observable $S_z(t)$, namely
  \begin{equation}
       S_{z}(t)=\frac{1}{J}\langle\psi_0|J_z(t)|\psi_0\rangle.
   \end{equation}
Here the initial states are chosen as the coherent states studied in the previous section. To see them clear,  we marked our initial states in Fig.\ref{psren} (e) and Fig.\ref{psren}(h) by using the diamond, circle and star points, respectively. We display our numerical results in Fig.\ref{Szt}.

The first row of Fig.\ref{Szt} illustrates the dynamics in the integrable regime. When the initial states originate from short closed orbits, $S_z(t)$ exhibits persistently oscillating behavior as shown in Fig.\ref{Szt}(a). However, if the initial states are associated with long periodic orbits or unstable fixed points (Fig.\ref{Szt} (b) and (c)), the oscillations eventually decay.
In the chaotic regimes, for the majority of initial states, as depicted in Fig.\ref{Szt} (d), $S_z(t)$ rapidly decays to a stationary state.When the initial state is selected at the unstable fixed point, $S_z(t)$ decays more slowly in an oscillatory manner, as depicted in Fig. \ref{Szt} (e). Interestingly, it exhibits a 4-periodic oscillation behavior, consistent with our discussion in Appendix~\ref{4p fixed point}. Conversely, within the integrable islands, the dynamics of $S_z$ remains frozen at the initial states, exhibiting a self-trapping behavior, as shown in Fig.\ref{Szt} (f).

\begin{figure}[h]
    \centering
    \includegraphics[width=0.5\textwidth,height=0.26\textwidth]{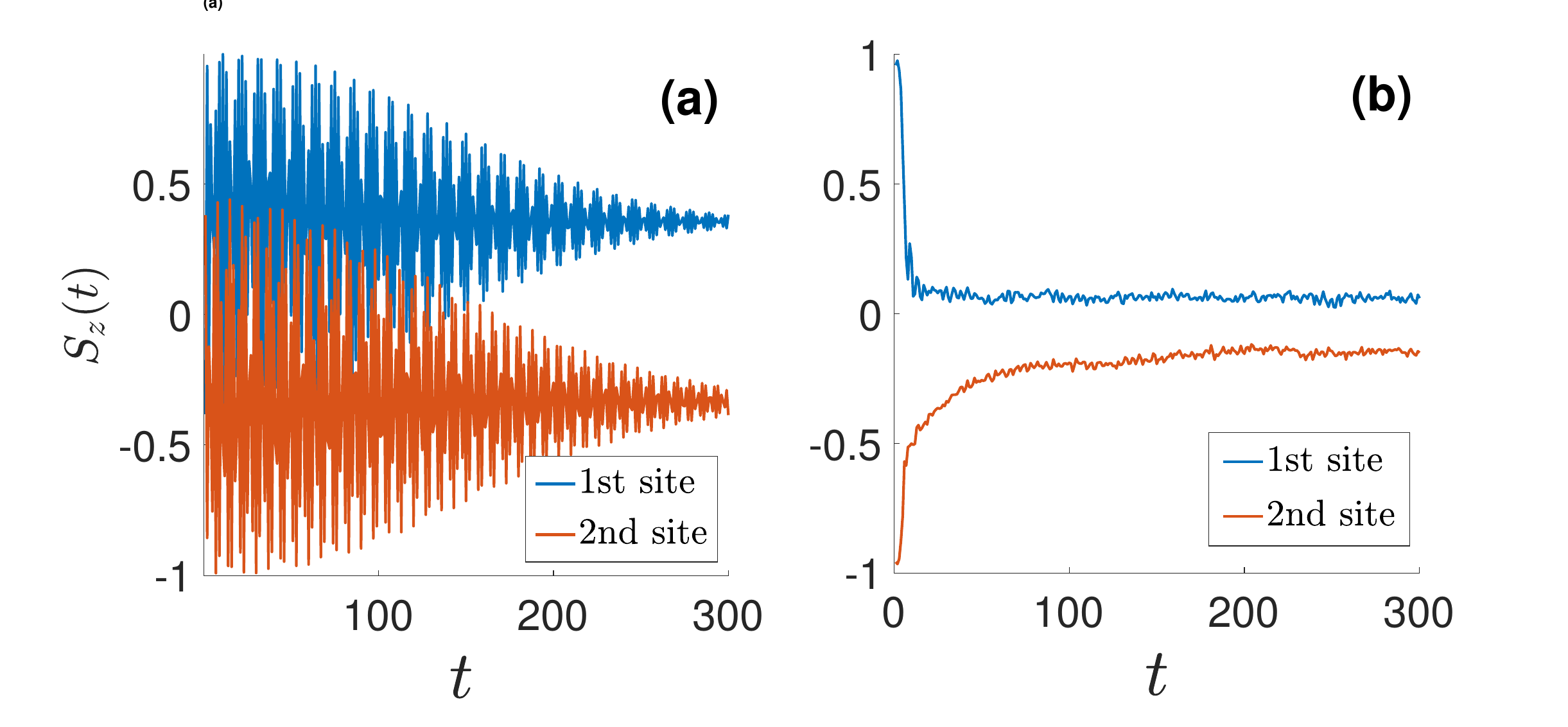}
    \caption{Dynamics of $S_z$ from different initial product states. (a) is in the regular regime with parameters $\mu=3$ and $k=1$. (b) is in the chaotic regime with parameters $\mu=3$ and $k=8$. Here we fix $J=800$.}
    \label{bsz}
\end{figure}

Next we demonstrate the dynamics of $S_z(t)$ by choosing distinct initial states as product states of bosons: one with bosons occupying solely on the first site, another  solely on the second site. As depicted in Fig. \ref{bsz}, the initial states of single-site occupation are far from equilibrium for both the integrable and chaotic cases, but they reach equilibrium in different ways. While $S_z(t)$ exhibits obviously oscillating behavior in the regular regime, it rapidly decays to a stationary state in the chaotic regime.

\subsection{Dynamics of OTOC}
\begin{figure*}
  \centering
  \includegraphics[width=1\textwidth,height=0.48\textwidth]{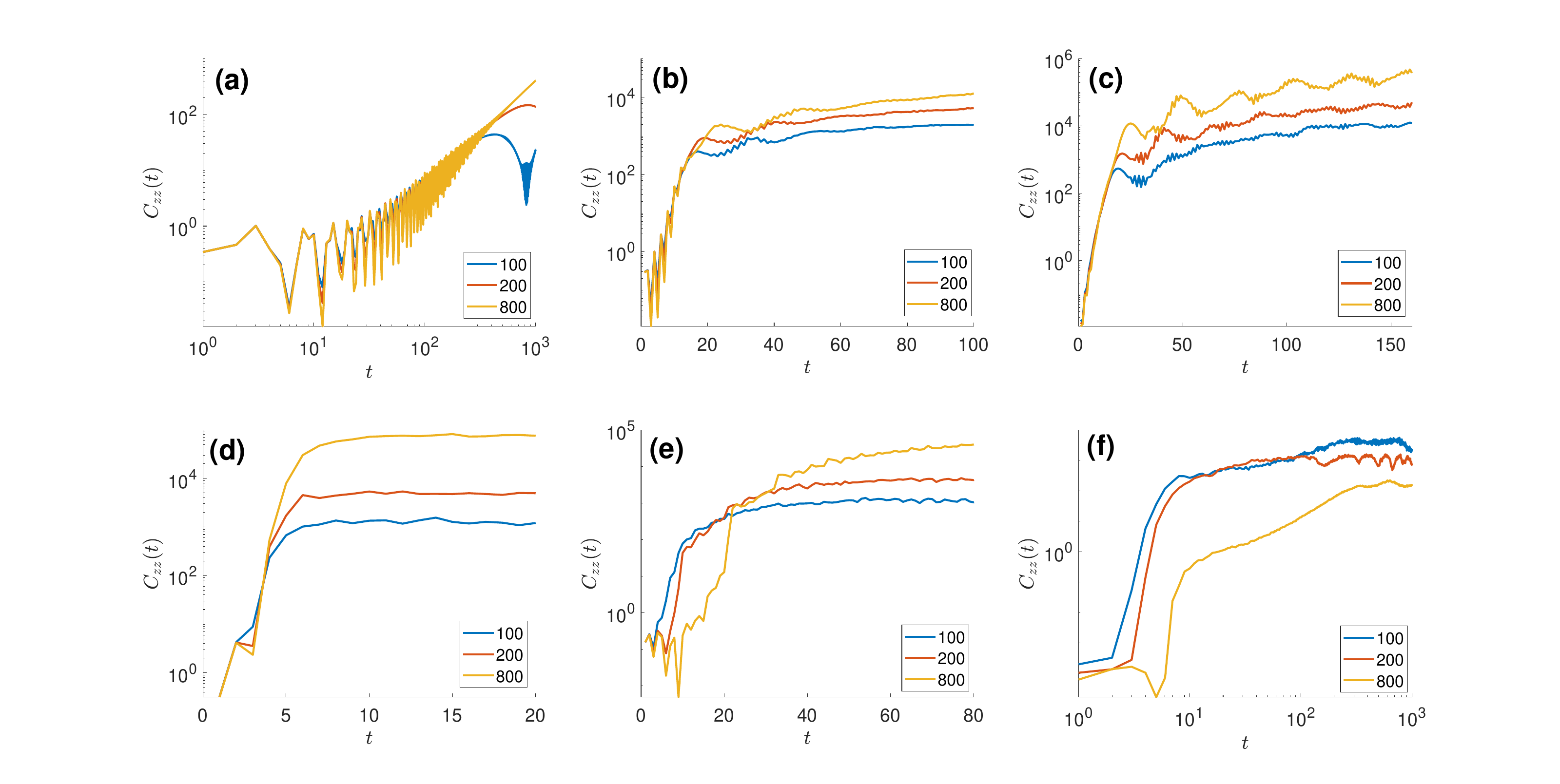}
  \caption{OTOC initialized from different initial coherent states. The first row: (a)-(c) are in the regular regime with parameters $\mu=3$ and $k=1$. From left to right, the initial states correspond to diamond, circle and star points marked in Fig.\ref{psren}(e). The second row: (d)-(f) are in the chaotic regime with parameters $\mu=3$ and $k=8$.
  From left to right, the  initial states correspond to  diamond, circle and star points marked in Fig.\ref{psren} (h). Here $J=100,200$ and $800$, respectively. }
  \label{otoc}
\end{figure*}

   The OTOC, as a new way to understand and diagnose quantum chaos, has been discussed from many different fields in recent years \cite{Maldacena,Swingle,Swingle-pra,OTOC2017,OTOC2022,OTOC19,Xu}. It is defined as
   \begin{equation}
   C(t)=-\left\langle [\hat{W}(t),\hat{V}]^2\right\rangle,
    \end{equation}
    where $\left\langle \cdots \right\rangle$ denotes the expectation values, $\hat{W}$ and $\hat{V}$ are two operators, and $\hat{W}(t)=U^{\dagger}W U$. Note that $C(t) \geq 0$ if $\hat{W}$ and $\hat{V}$ are Hermitian operators.  Analogy to the classical Lyaponouv exponents, OTOC is expected to have the exponential growth at early time in quantum chaotic systems. While recently some works found for integrable model the OTOC also can have exponential growth if the initial states near the non-stable fixed point in classical limit \cite{Richter,CaoX,Galitski2020,Richter2023,Cameo,Kidd2021}. Here we study the square commutator between two $J_z$ operator, namely
    \begin{equation}
       C_{zz}(t)=-\left(\frac{1}{J}\right)^2\langle\psi_0|\left[J_z(t),J_z\right]^2|\psi_0\rangle,
   \end{equation}
    where we still choose the coherent state given by \eqref{coherentstates} as the initial state $|\psi_0\rangle$.
    As we will illustrate in Fig.\ref{otoc}, in both regular and chaotic regimes, the growth behaviour of OTOC strongly depends on its initial states. Based on the Poincar\'{e} section and multifractal dimensions of coherent states, we select separately three representative initial states for regular and chaotic cases, respectively, as marked in Fig.\ref{psren} (e) and Fig.\ref{psren} (h) by using diamond, circle and star points.

\begin{figure}[h]
    \centering
    \includegraphics[width=0.5\textwidth,height=0.26\textwidth]{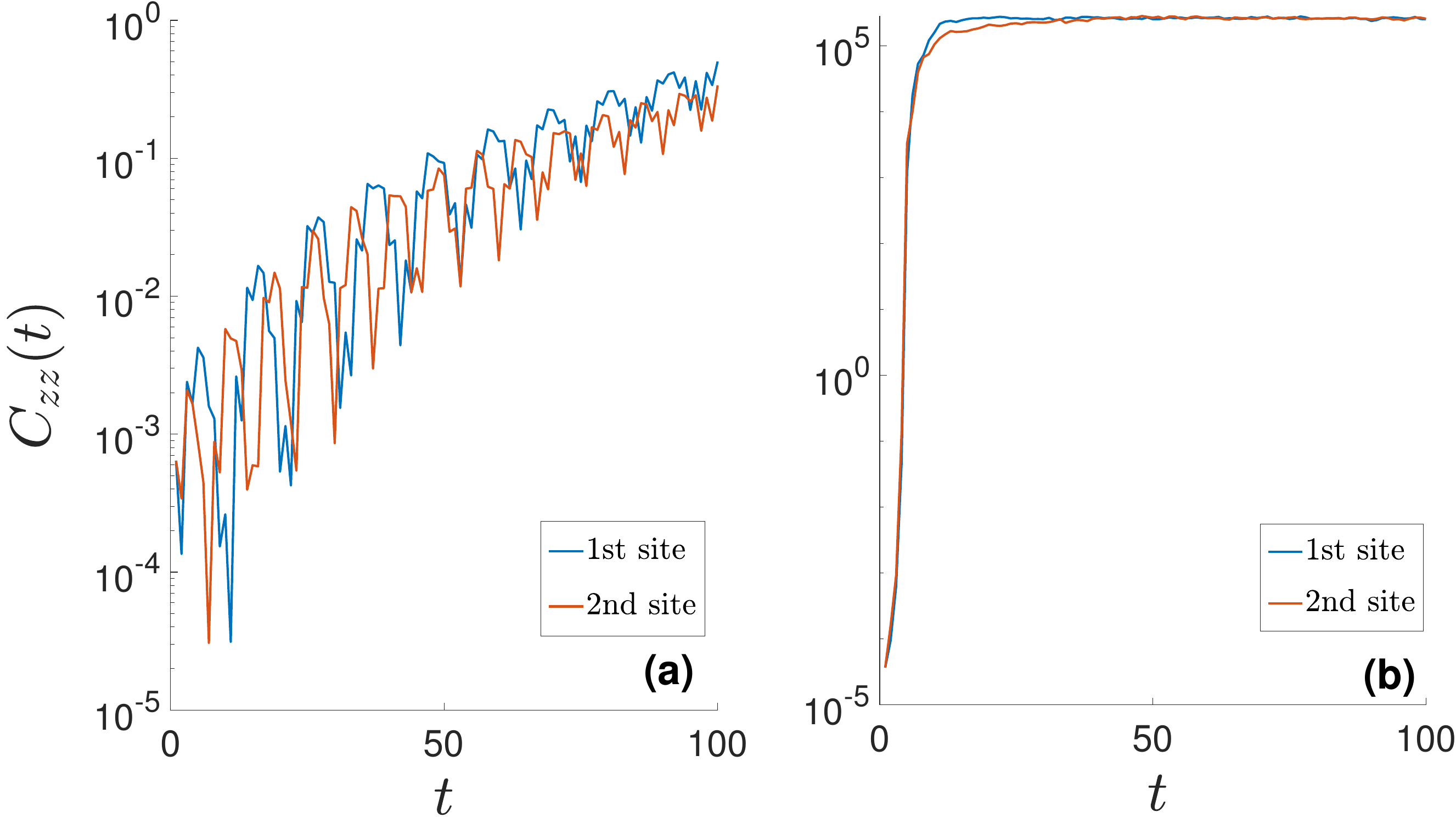}
    \caption{Dynamics of OTOC from different initial product states. (a) is in the regular regime with parameters $\mu=3$ and $k=1$. (b) is in the chaotic regime with parameters $\mu=3$ and $k=8$. Here we fix $J=800$.}
    \label{botoc}
\end{figure}

In the regular regime with $k=1$, the diamond point is situated on short closed periodic orbits near the stable fixed point, exhibiting very small R\'{e}nyi entropy. The OTOC from this point, as depicted in Fig.\ref{otoc}(a), exhibits a slow linear growth, and the saturation values are remarkably small. In contrast, the circle point lies on a long periodic orbit in the Poincar\'{e} section, featuring higher R\'{e}nyi entropy compared to its surrounding points. This orbit connects to unstable fixed points in the Poincar\'{e} section. The OTOC from this circle point, illustrated in Fig. \ref{otoc} (b), grows exponentially at early times, reaching a larger saturation value. When the star point is situated near the unstable fixed point in the Poincar\'{e} section, the OTOC growth is clearly exponential at early times, as shown in Fig.\ref{otoc} (c), resembling the chaos case discussed below. Additionally, its saturation values are larger than those of the former point, reaching a level comparable to the chaotic case.

In the chaotic regime with $k=8$, the diamond point exhibits a large R\'{e}nyi entropy, corresponding to a randomly chosen point in the Poincar\'{e} section, which represents a majority of points in the phase space. The OTOC evolving from this initial state, as depicted in Fig.\ref{otoc} (d), exhibits standard exponential growth at early times and quickly reaches saturation. The circle point is situated at an isolated unstable fixed point, surrounded by random trajectories. The OTOC evolving from this circle point, as illustrated in Fig.\ref{otoc} (e), also exhibits exponential growth, but grows much slower than the case from the diamond point. Finally, the star point is located in deep integrable islands, featuring minimal R\'{e}nyi entropy. The OTOC evolving from this star point, as shown in Fig.\ref{otoc} (f),  no longer grows exponentially, and its saturation value is much smaller than the values  in Figs.\ref{otoc} (d) and (e), reaching the same level as the regular case. With the increase in $J$, the saturation value is further suppressed.

We also demonstrate the dynamics of OTOC with the initial state chosen as distinct product states of bosons: one with bosons occupying solely on the first site, another solely on the second site. As illustrated in Fig. \ref{botoc}, while the OTOC exhibits a slow linear growth in the regular regime of $k=1$,  it exhibits exponential growth at early times in the the chaotic regime of $k=8$ for the initial state either on the first or second site.

\begin{figure}
  \centering
  \includegraphics[width=0.4\textwidth,height=0.6\textwidth]{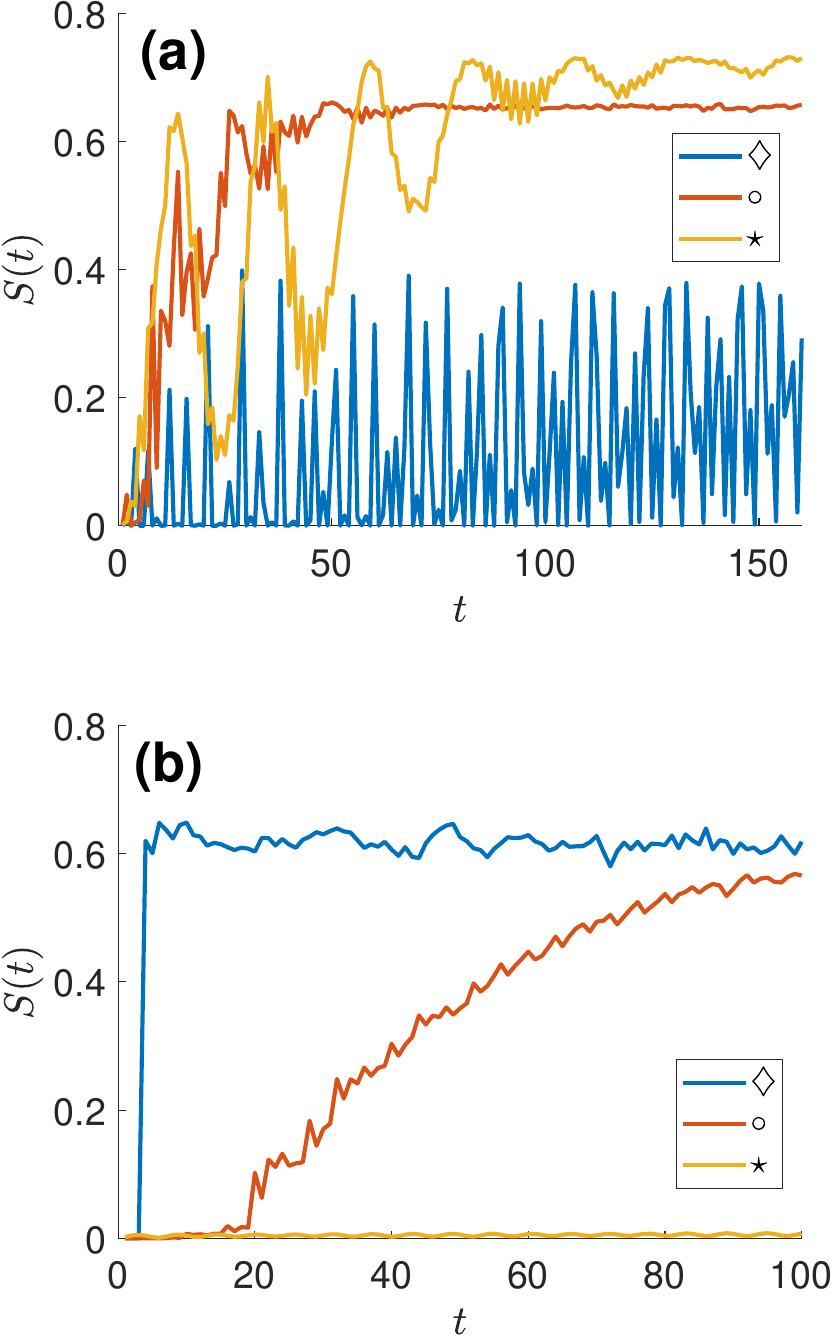}
  \caption{Entanglement entropy versus $t$ for different initial coherent states. (a) is in the regular regime with parameters $\mu=3$ and $k=1$. Three initial states
  correspond to diamond, circle and star points in Fig.\ref{psren} (e), respectively. (b) is in the chaotic regime with parameters $\mu=3$ and $k=8$. Three initial states correspond to diamond,
  circle and star points in Fig.\ref{psren} (h), respectively. Here we take $J=400$.}
  \label{entan}
\end{figure}

\subsection{Dynamics of entanglement entropy}
The dynamics of entanglement entropy  has been applied to detect chaos in quantum systems \cite{WangXG,znidaric08,Huse13,bianchijhep,bianchipra}. A large spin system can be effectively represented as a many-qubit system, such as in our model and the LMG model, both of which can be viewed as systems composed of
$J$ qubits. Employing the decomposition method from \cite{WangXG,Vidal05,Gisin}, we calculate the von Neumann entanglement entropy of the reduced density matrix $\rho_s$
after tracing out  $s'=J-s$ qubits: $S_E=-\text{Tr}{\rho_s \ln{\rho_s}}$. In the following calculation, we consider $s=2$. Using the same initial states as utilized in the dynamics of OTOC, we explore the dynamical evolution of the entanglement entropy.

Similar to the behaviors observed in the OTOC for different initial states, in the
regular regime with $k=1$, as depicted in Fig.\ref{entan} (a), the entanglement originating from a small closed orbit exhibits persistent oscillations. Meanwhile, when initialized from a long periodic orbit or an unstable fixed point, the entanglement entropy grows with oscillations and reaches a higher saturation value. In the chaotic regime with $k=8$, as illustrated in Fig.\ref{entan} (b), the entanglement entropy experiences rapid growth, swiftly approaching saturation when initialized from the point where the OTOC exhibits standard exponential growth. In contrast, for the case of an unstable fixed point, akin to the OTOC dynamics, the entanglement entropy reaches saturation at a more gradual pace compared to the majority of cases. Finally, when initialized from deep integrable islands, the entanglement entropy remains stagnant and hovers around zero.

\section{Summary and outlook}
\label{sec:outlook}
In summary, we systematically study a kicked Bose Hubbard dimer, which can be effectively described by a kicked LMG model, and uncover signatures of chaos in the system from both statistical and dynamical aspects. The diagnosis of spectral statistics of Floquet operator reveals the periodical dependence of $\mu$ and the existence of transition from regular to chaotic dynamics with the increase in $k$. Furthermore, insights from  semiclassical dynamics and analysis of phase space localization measure suggest that the system may
display fruitful dynamical phenomena in various parameter regimes with different initial states, which are confirmed by studying the dynamics of local operators, OTOC and entanglement entropy.  Our results exhibit the richness of dynamics in the kicked Bose-Hubbard dimer.

Our results build on the kicked Bose-Hubbard dimer and are important complementary to previous studies based on the kicked top model.  Due to the good controllability of cold atomic platforms, our model is principally realizable in current cold atomic experiments, and thus provides an alternative protocol to study the interesting dynamical phenomena beyond the standard kicked top model. Owing to the absence of parity symmetry, it would be interesting to explore what new features beyond the quantum kicked top can be found in the future work.

\begin{acknowledgments}
We thank L. B. Fu for helpful discussions. The financial supports from the National Natural Science Foundation of China (Grant No.  12174436
and T2121001), and Strategic Priority Research Program of the Chinese Academy of Sciences (Grant No. XDB33000000) are gratefully acknowledged.

\end{acknowledgments}

\appendix
\section{Poincar\'{e} section on sphere topology}
\label{sphere PS}
\begin{figure}[h]
    \centering
    \includegraphics[width=0.5\textwidth,height=0.42\textwidth]{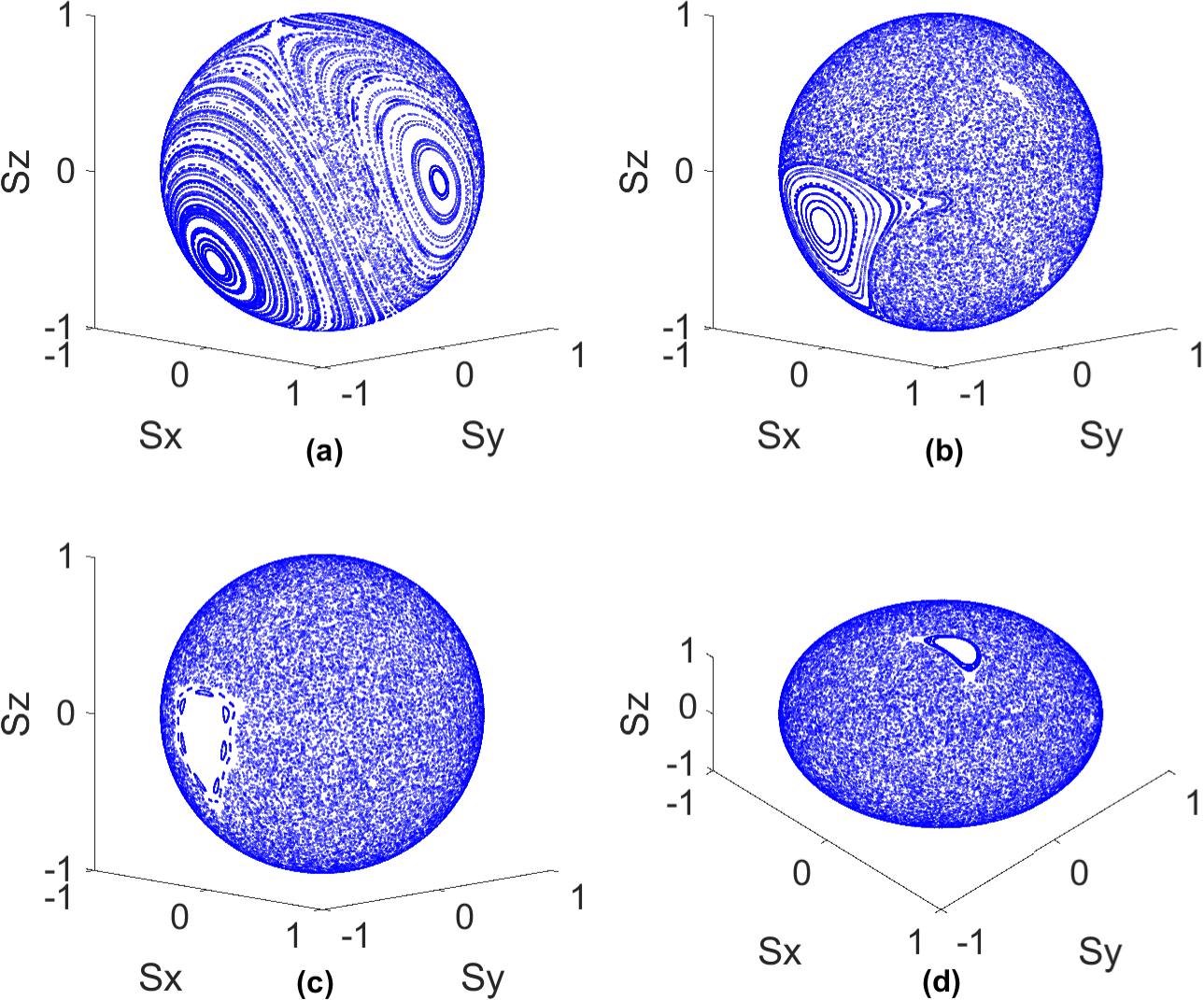}
    \caption{(a)-(d): Poincar\'{e} section of the classic equation of motions on the sphere phase space.Each trajectory is evolved for 400 kicks. The parameter $\mu$ is fixed at 3, and $k=1,3,5$, and $8$ from (a) to (d), respectively.}
    \label{pss}
\end{figure}

Here we provide the Poincar\'{e} section on sphere, which is a true manifold for phase space of a spin model. As shown in Fig.\ref{pss}, we take the parameter $\mu$ fixed at 3, and $k=1,3,5$, and $8$ from (a) to (d), corresponding to Fig.4 (a)-(d), respectively.

\section{Analysis of properties of low-R\'{e}nyi-entropy points in chaotic regime}
\label{4p fixed point}
\begin{figure}[h]
    \centering
    \includegraphics[width=0.5\textwidth,height=0.5\textwidth]{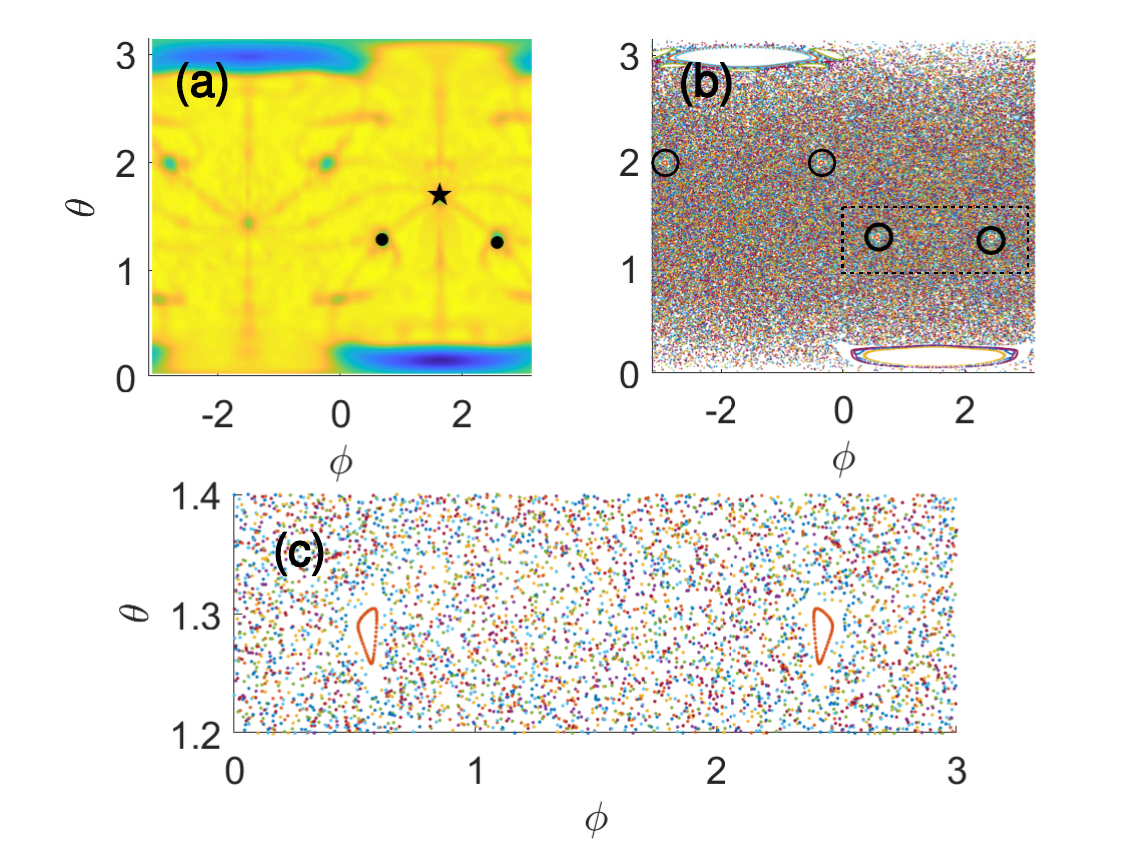}
    \caption{(a): Multifractal dimension of coherent state, (b): Poincar\'{e} section. The parameters are $\mu=3$ and $k=8$. The black points marked in the figure correspond to the fixed points studied in the appendix~\ref{4p fixed point}. (c): The Poincar\'{e} section corresponding to the region delineated by the dotted line in Fig.\ref{fixpoint} (b).}
    \label{fixpoint}
\end{figure}

\begin{figure}[h]
    \centering
    \includegraphics[width=0.4\textwidth,height=0.3\textwidth]{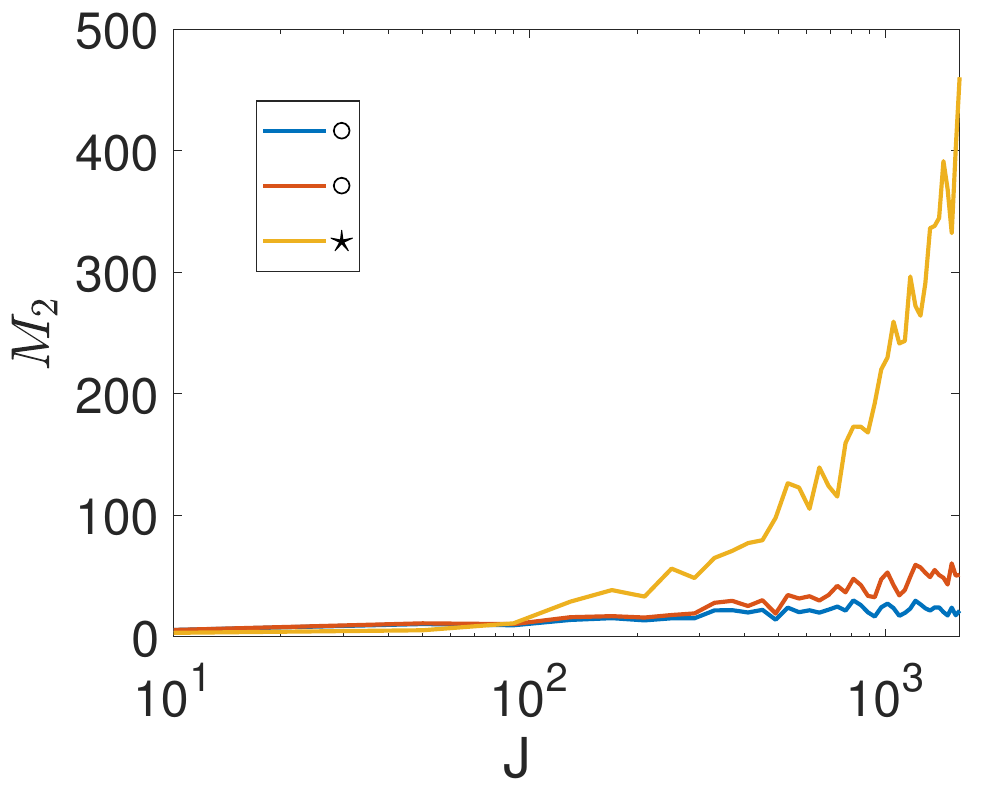}
    \caption{Size scaling of participation number $M_2$ of three coherent states marked in Fig. \ref{fixpoint}(a). Here $J$ is total spin number, and the Hilbert space dimension $N=2J+1$.}
    \label{M2s}
\end{figure}

\begin{figure}[h]
    \centering
    \includegraphics[width=0.52\textwidth,height=0.42\textwidth]{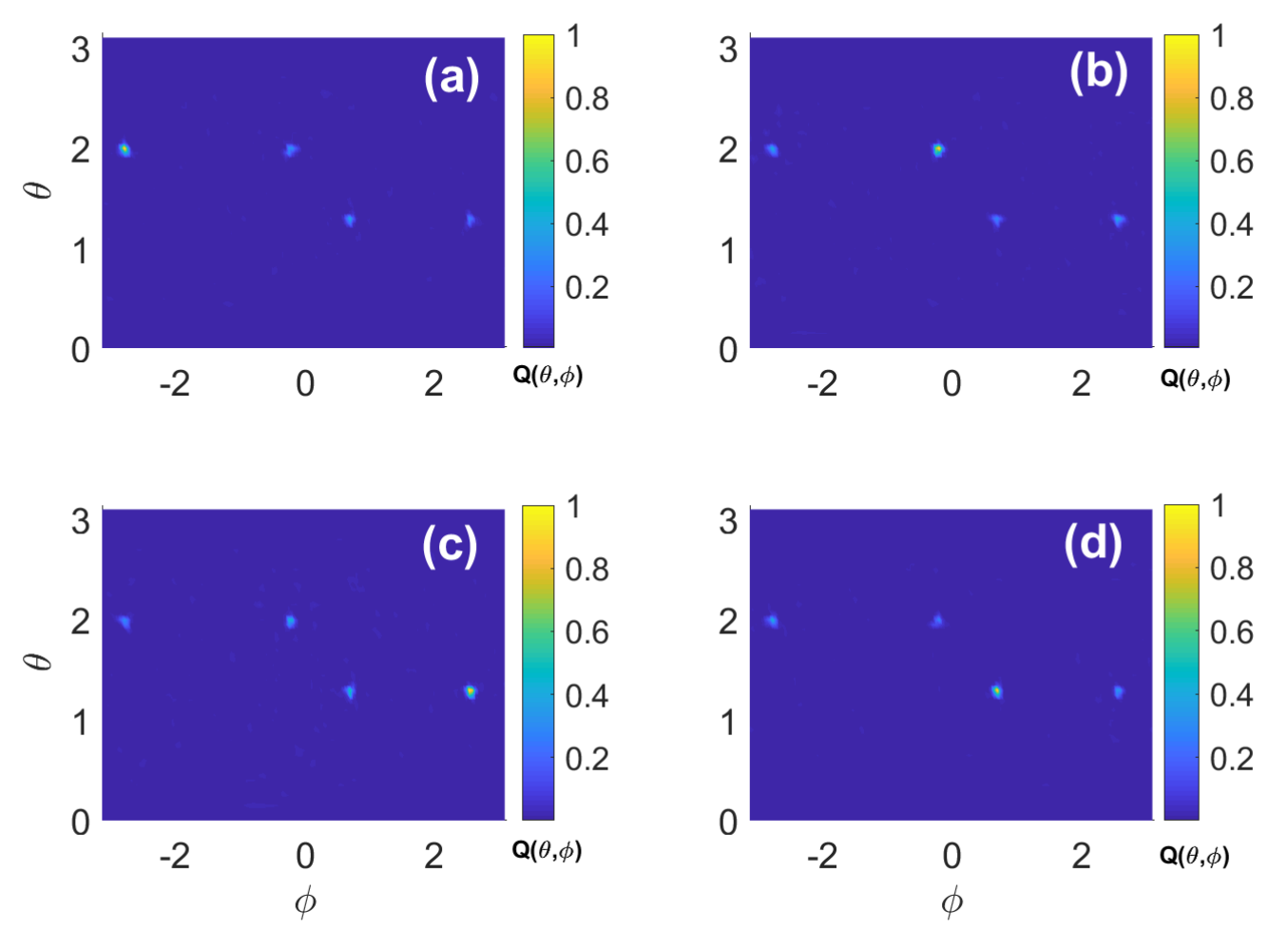}
    \caption{The Husimi function of the coherent state at different time, starting from the left circular point in Fig. \ref{fixpoint} (a). $T=1000, 1001, 1002$, and $1003$ correspond to panels (a) to (d), respectively.}
    \label{4p}
\end{figure}

In sec.\ref{sec:coherntrenyi}, we observe that there are still some  points with low R\'{e}nyi entropy in the bulk of phase space even when the model enters into the chaotic regime of $k=8$, as illustrated in Fig.\ref{psren}.(h). Here we show these points belonging to different classes using the size scaling of participation number \cite{BHDimer2013}. The participation number is closely associated with the R\'{e}nyi entropy $S_2$ of coherent states \ref{eq:renyi}, defined as $M_2=\frac{1}{\sum_{i=1}^{\mathcal{N}}|c_{i}|^{4}} = \exp{S_2}$. We specifically examine the participation number of the points depicted in Fig.\ref{fixpoint} (a). As illustrated in Fig.\ref{M2s}, the participation number of the star point increases with the dimension of the Hilbert space, while the circular points almost keep invariant despite of changes in the Hilbert space dimension. These two circular points correspond to hidden fixed points in the Poincar'{e} section marked in Fig. \ref{fixpoint} (b), whereas the star point does not correspond any fixed point in classical phase space. To clarify, in Fig. \ref{fixpoint} (c), we display magnifications of the Poincar\'{e} section around the two circular points marked by the dotted line in Fig. \ref{fixpoint} (b). We also exhibit the long time dynamics of Husimi function of these fixed points in Fig.\ref{4p}. Initiating with the coherent state localized at the left circular point in Fig.\ref{fixpoint}.(a) (circle point in Fig.4(h)), we set $J=1000$ and examine the behavior at $T=1000, 1001, 1002, 1003$. Throughout these time instances, the Husimi function consistently remains localized predominantly at distinct fixed points marked in Fig. \ref{fixpoint} (b). The dynamics for the initial state chosen at the fixed point displays a 4-periodic oscillating behavior.

\begin{thebibliography}{}
\bibitem{LZ1} L. D. Landau, Zur Theorie der Energieubertragung II,  Phys. Z. Sowjetunion 2, 46 (1932).
\bibitem{LZ2} G. Zener,  Non-adiabatic crossing of energy levels, Proc. R. Soc. London A {\bf 137}, 696 (1932).

\bibitem{WuB} B. Wu and Q. Niu, Nonlinear Landau-Zener tunneling, Phys. Rev. A {\bf 61}, 023402 (2000).

\bibitem{DW1} G. J. Milburn, J. Corney, E. M. Wright, and D. F. Walls, Quantum dynamics of an atomic Bose-Einstein condensate in a double-well potential,
Phys. Rev. A {\bf 55}, 4318 (1997).

\bibitem{DW2} A. Smerzi, S. Fantoni, S. Giovanazzi, and S. R. Shenoy, Quantum Coherent Atomic Tunneling between Two Trapped Bose-Einstein Condensates,
Phys. Rev. Lett. {\bf 79}, 4950 (1997).

\bibitem{Smerzi1999} S. Raghavan, A. Smerzi, S. Fantoni, and S. R. Shenoy, Coherent oscillations between two weakly coupled Bose-Einstein condensates: Josephson effects, $\pi$ oscillations, and macroscopic quantum self-trapping, Phys. Rev. A {\bf 59}, 620 (1999).
\bibitem{Smerzi2001} J. R. Anglin, P. Drummond, and A. Smerzi, Exact quantum phase model for mesoscopic Josephson junctions,
Phys. Rev. A {\bf 64}, 063605 (2001).
\bibitem{FuLB} G.-F. Wang, L.-B. Fu, and J. Liu, Periodic modulation effect on self-trapping of two weakly coupled Bose-Einstein condensates,
Phys. Rev. A {\bf 73}, 013619 (2006).

\bibitem{Strzys} M P Strzys, E. M. Graefe, and H. J. Korsch, Kicked Bose Hubbard systems and kicked tops-destruction and stimulation of tunneling, New J. Phys. {\bf 10}, 013024 (2008).

\bibitem{BHDimer2013} C. Khripkov, D. Cohen, and A. Vardi, Coherence dynamics of kicked Bose-Hubbard dimers: Interferometric signatures of chaos,
Phys. Rev. E {\bf 87}, 012910 (2013).

\bibitem{Kidd} R. A. Kidd, M. K. Olsen, and J. F. Corney, Quantum chaos in a Bose-Hubbard dimer with modulated tunneling,
Phys. Rev. A {\bf 100}, 013625 (2019).

\bibitem{pencil1}T. Dittrich, and S. P. Mart\'{i}nez, Toppling Pencils - Macroscopic Randomness from Microscopic Fluctuations. Entropy {\bf 2020}, 22, 1046 (2020).

\bibitem{pencil2}S. L. Choudhury and F. Grossmann, Quantum approach to the thermalization of the toppling pencil interacting with a finite bath,
Phys. Rev. A {\bf 105}, 022201 (2022).


\bibitem{Leggett}  A. J. Leggett, Bose-Einstein condensation in the alkali gases: Some fundamental concepts, Rev. Mod. Phys. 73, 307 (2001).
\bibitem{Korsch} E. M. Graefe and H. J. Korsch, Semiclassical quantization of an N-particle Bose-Hubbard model
Phys. Rev. A {\bf 76}, 032116 (2007).

\bibitem{Chuchem} M. Chuchem, K. Smith-Mannschott, M. Hiller, T. Kottos, A. Vardi, and D. Cohen, Quantum dynamics in the bosonic Josephson junction,
Phys. Rev. A {\bf 82}, 053617 (2010).

\bibitem{Links} J. Links and H.-Q. Zhou, Exact Form Factors for the Josephson Tunneling Current and Relative Particle Number Fluctuations in a Model of Two Coupled Bose Einstein Condensates, Lett. Math. Phys. 60, 275 (2002).

\bibitem{ZhouHQ}  Huan-Qiang Zhou, Jon Links, R. H McKenzie and Xi-Wen Guan, Exact results for a tunnel-coupled pair of trapped Bose Einstein condensates, J. Phys. A: Math. Gen. 36 L113 (2003).



\bibitem{Albiez} M. Albiez, R. Gati, J. F\"{o}lling, S. Hunsmann, M. Cristiani, and M. K. Oberthaler, Direct Observation of Tunneling and Nonlinear Self-Trapping in a Single Bosonic Josephson Junction, Phys. Rev. Lett. {\bf 95}, 010402 (2005).

\bibitem{Levy} S. Levy, E. Lahoud, I. Shomroni, and J. Steinhauer, The a.c. and d.c. Josephson effects in a Bose-Einstein condensate, Nature
(London) {\bf 449}, 579 (2007).

\bibitem{Exp-kicktop}J. Tomkovic, W. Muessel, H. Strobel, S. L\"{o}ck, P. Schlagheck, R. Ketzmerick, and M. K. Oberthaler, Experimental observation of the Poincar\'{e}-Birkhoff scenario in a driven many-body quantum system, Phys. Rev. A 95, 011602(R) (2017).

\bibitem{Exp-kicktop2} S. Chaudhury, A. Smith, B. E. Anderson, S. Ghose and P. S. Jessen, Quantum signatures of chaos in a kicked top, Nature, {\bf 461}, 768 (2009).

\bibitem{Tonel} A. P. Tonel, J. Links, A. Foerster,  Quantum dynamics of a model for two Josephson-coupled Bose--Einstein condensates, J, Phys. A {\bf 38}, 1235 (2005).



\bibitem{Kicktop} F. Haake, M. Ku\'{s}, and R. Scharf, Classical and quantum chaos for a kicked top, Z. Phys. B: Condens. Matter 65, 381 (1987).


\bibitem{Fox} R. F. Fox and T. C. Elston, Chaos and a quantum-classical correspondence in the kicked top,
Phys. Rev. E {\bf 50}, 2553 (1994).

\bibitem{Kicktop1997}V. Constantoudis and N. Theodorakopoulos, Lyapunov exponent, stretching numbers, and islands of stability of the kicked top. Phys. Rev. E {\bf 56}, 5189 (1997).


\bibitem{WangXG} X. Wang, S. Ghose, B. C. Sanders, and B. Hu, Entanglement as a signature of quantum chaos,
Phys. Rev. E {\bf 70}, 016217 (2004).

\bibitem{Ghose} S. Ghose, R. Stock, P. Jessen, R. Lal, and A. Silberfarb, Chaos, entanglement, and decoherence in the quantum kicked top, Phys. Rev. A {\bf 78}, 042318 (2008).

\bibitem{Lombardi} M. Lombardi and A. Matzkin, Entanglement and chaos in the kicked top,
Phys. Rev. E {\bf 83}, 016207 (2011).

\bibitem{Dorgra} S. Dogra, V. Madhok, and A. Lakshminarayan, Quantum signatures of chaos, thermalization, and tunneling in the exactly solvable few-body kicked top,
Phys. Rev. E {\bf 99}, 062217 (2019).

\bibitem{Kidd2020} R. A. Kidd, A. Safavi-Naini, and J. F. Corney, Thermalization in a Bose-Hubbard dimer with modulated tunneling,
Phys. Rev. A {\bf 102}, 023330 (2020).

\bibitem{KickPtop}Manuel H. Mu$\tilde{n}$oz-Arias, Pablo M. Poggi, and Ivan H. Deutsch,Nonlinear dynamics and quantum chaos of a family of kicked
p-spin models,
Phys. Rev. E {\bf 103}, 052212

\bibitem{lerosepra} A. Lerose and S. Pappalardi, Bridging entanglement dynamics and chaos in semiclassical systems,  Phys. Rev. A, {\bf 102}, 032404 (2020).

\bibitem{WangQ2021}Q. Wang and M. Robnik, Multifractality in Quasienergy Space of Coherent States as a Signature of Quantum Chaos
Signature of Quantum Chaos, Entropy, {\bf 23}, 1347 (2021).

\bibitem{WangQ2023}Q. Wang and M. Robnik, Statistics of phase space localization measures and quantum chaos in the kicked top model, Phys. Rev. E {\bf 107}, 054213 (2023).


\bibitem{Berry} M. Berry, Quantum chaology, not quantum chaos, Phys. Scr. {\bf 40}, 335 (1989).

\bibitem{Haake} 
F. Haake, Quantum Signatures of Chaos, 3rd ed. (Springer, Berlin, 2010).

\bibitem{WuB2021}Z. Wang, Y. Wang, and B. Wu, Quantum chaos and physical distance between quantum states,
Phys. Rev. E {\bf 103}, 042209 (2021).


\bibitem{BGS} O. Bohigas, M.J. Giannoni and C. Schmit, Characterization of Chaotic Quantum Spectra and Universality of Level Fluctuation Laws. Phys. Rev. Lett. {\bf 52}, 1 (1984).

\bibitem{Berry-1977spectrum} M. V. Berry and M. Tabor, Level clustering in the regular spectrum. Proc. R. Soc. A {\bf 356}, 375 (1977).

\bibitem{Huse} V. Oganesyan and D. A. Huse, Localization of interacting fermions at high temperature. Phys. Rev. B {\bf 75}, 155111 (2007).

\bibitem{Atas} Y. Y. Atas, E. Bogomolny, O. Giraud and G. Roux, Distribution of the Ratio of Consecutive Level Spacings in Random Matrix Ensembles, Phys. Rev. Lett. {\bf 110}, 084101 (2013).

\bibitem{Guhr}T. Guhr, A. Mueller-Gr\"{o}eling, and H. A. Weidenm ller, Random matrix theories in quantum physics: Common concepts,
Phys. Rep. {\bf 299}, 189 (1998).

\bibitem{PhysRep}F. M. Izrailev, Simple models of quantum chaos: Spectrum and eigenfunctions,  Phys. Rep. {\bf 196}, 299 (1990).

\bibitem{Berry1977} M.V. Berry, Regular and irregular semiclassical wavefunctions. J. Phys. A  {\bf 10}, 2083 (1977).

\bibitem{Kus} M. Kus, J. Mostowski and F. Haake, Universality of eigenvector statistics of kicked tops of different symmetries, J. Phys. A: Math. Gen. 21 L1073 (1988).

\bibitem{Zyczkowski} K. Zyczkowski, Indicators of quantum chaos based on eigenvector statistics,  J. Phys. A: Math. Gen. {\bf 23}, 4427 (1990).

\bibitem{Haake1990} F. Haake and K. Zyczkowski, Random-matrix theory and eigenmodes of dynamical systems. Phys. Rev. A {\bf 42}, 1013 (1990).

\bibitem{znidaric08}Marko Znidaric, Tomaz Prosen, and Peter Prelovsek,Many-body localization in the Heisenberg xxz magnet in a random field,
Phys. Rev. B {\bf 77}, 064426 (2008).

\bibitem{Huse13}Hyungwon Kim and David A. Huse, Ballistic Spreading of Entanglement in a Diffusive Nonintegrable System,
Phys. Rev. Lett. {\bf 111}, 127205 (2013)

\bibitem{bianchijhep} Bianchi, E., Hackl, L.  Yokomizo, N. Linear growth of the entanglement entropy and the Kolmogorov-Sinai rate. J. High Energ. Phys. {\bf 2018}, 25 (2018).

\bibitem{bianchipra} Lucas Hackl, Eugenio Bianchi, Ranjan Modak, and Marcos Rigol, Entanglement production in bosonic systems: Linear and logarithmic growth,
Phys. Rev. A {\bf 97}, 032321 (2018)

\bibitem{Nahum}Adam Nahum, Jonathan Ruhman, Sagar Vijay, and Jeongwan Haah,Quantum Entanglement Growth under Random Unitary Dynamics,
Phys. Rev. X {\bf 7}, 031016

\bibitem{Fazio} Silvia Pappalardi, Angelo Russomanno, Bojan Zunkovic, Fernando Iemini, Alessandro Silva, and Rosario Fazio, Scrambling and entanglement spreading in long-range spin chains,
Phys. Rev. B {\bf 98}, 134303 (2018).

\bibitem{Larkin} Larkin, Anatoly I., and Yu N. Ovchinnikov. "Quasiclassical method in the theory of superconductivity." Sov Phys JETP 28.6 (1969): 1200.

\bibitem{Kitaev}A. Kitaev, Hidden correlations in the hawking radiation and thermal noise, talks given at The Fundamental Physics Prize Symposium, 10 November 2014, and at The KITP, 12 February 2015.

\bibitem{Maldacena} J. Maldacena, S. H. Shenker, and D. Stanford, A bound on
chaos, J. High Energy Phys. 08 (2016) 106.

\bibitem{Swingle-pra} B. Swingle, G. Bentsen, M. Schleier-Smith, and P. Hayden,
Measuring the scrambling of quantum information, Phys. Rev.A {\bf 94}, 040302(R) (2016).

\bibitem{Swingle}  B. Swingle, Unscrambling the physics of out-of-time-order correlators, Nat. Phys. {\bf 14}, 988 (2018).

\bibitem{Xu} S. Xu, and B. Swingle, Scrambling dynamics and out-of-time ordered correlators in quantum many-body systems: a tutorial, arXiv:2202.07060.

\bibitem{OTOC2017} E. B. Rozenbaum, S. Ganeshan, and V. Galitski, Lyapunov Exponent and Out-of-Time-Ordered Correlator's Growth Rate in a Chaotic System, Phys. Rev. Lett. {\bf 118}, 086801 (2017).

\bibitem{OTOC2022} M. Zonnios, J. Levinsen, M. M. Parish, F. A. Pollock, and K. Modi, Signatures of Quantum Chaos in an Out-of-Time-Order Tensor, Phys. Rev. Lett. {\bf 128}, 150601 (2022).

\bibitem{OTOC19} R. J. Lewis-Swan, A. Safavi-Naini, J. J. Bollinger, and A. M. Rey, Unifying scrambling, thermalization and entanglement through measurement of fidelity out-of-time-order correlators in the Dicke model, Nat. Commun. {\bf 10}, 1581 (2019).

\bibitem{JPD} Sreeram PG, V. Madhok, and A. Lakshminarayan, Out-of-time-ordered correlators and the Loschmidt echo in the quantum
kicked top: how low can we go? J. Phys. D {\bf 54}, 274004 (2021).

\bibitem{Trunin} D. A. Trunin, Quantum chaos without false positives, Phys. Rev. D {\bf 108}, L101703 (2023).

\bibitem{Richter} Q. Hummel, B. Geiger, J. D. Urbina, and K. Richter, Reversible Quantum Information Spreading in Many-Body Systems near Criticality,
Phys. Rev. Lett. {\bf 123}, 160401 (2019).

\bibitem{CaoX}T. Xu, T. Scaffidi, and X. Cao, Does Scrambling Equal Chaos? Phys. Rev. Lett. {\bf 124}, 140602 (2020).

\bibitem{Galitski2020}E. B. Rozenbaum, L. A. Bunimovich, and V. Galitski,
Early-Time Exponential Instabilities in Nonchaotic Quantum Systems, Phys. Rev. Lett. {\bf 125}, 014101 (2020)

\bibitem{Cameo}S. Pilatowsky-Cameo, J. Ch\'{a}vez-Carlos, M. A. Bastarrachea-Magnani, P. Str\'{a}nsk\'{y}, S. Lerma-Hern\'{a}ndez, L. F. Santos, and J. G. Hirsch, Positive quantum Lyapunov exponents in experimental systems with a regular classical limit, Phys. Rev. E {\bf 101}, 010202(R) (2020).

\bibitem{Kidd2021} R. A. Kidd, A. Safavi-Naini, and J. F. Corney, Saddle-point scrambling without thermalization, Phys. Rev. A {\bf 103}, 033304 (2021).

\bibitem{Richter2023} M. Steinhuber, P. Schlagheck, J. D. Urbina, and K. Richter, Dynamical transition from localized to uniform scrambling in locally hyperbolic systems,
Phys. Rev. E {\bf 108}, 024216 (2023).

\bibitem{LMG2017} A. Russomanno, F. Iemini, M. Dalmonte, and R. Fazio, Floquet time crystal in the Lipkin-Meshkov-Glick model,
Phys. Rev. B {\bf 95}, 214307 (2017).

\bibitem{Wigner} E. Wigner, On the quantum correction for thermodynamic equilibrium, Phys. Rev. {\bf 40}, 749 (1932).
\bibitem{Husimi} K. Husimi, Some Formal Properties of the Density Matrix, Proc. Phys. Math. Soc. Jpn. {\bf 22}, 264 (1940).
\bibitem{ShiKJ} S.-J. Chang and K.-J. Shi, Time evolution and eigenstates of a quantum iterative system, Phys. Rev. Lett. {\bf 55}, 269 (1985).
\bibitem{Takahashi} K. Takahashi, Wigner and Husimi Functions in Quantum Mechanics, J. Phys. Soc. Jpn. {\bf 55}, 762 (1986).



\bibitem{WMZhang} W. M. Zhang, D. H. Feng and R. Gilmore, Coherent states: Theory and some applications. Rev. Mod. Phys. {\bf 62}, 867 (1990).

\bibitem{Coherentstate}J. P. Gazeau, Coherent States in Quantum Optics; Wiley-VCH: Berlin, Germany, 2009.

\bibitem{SCS}J. M. Radcliffe, Some properties of coherent spin states. J. Phys. A {\bf 4}, 313 (1971).

\bibitem{SCS2} G. M. Ariano, L. R. Evangelista and M. Saraceno, Classical and quantum structures in the kicked-top model. Phys. Rev. A {\bf 45}, 3646 (1992).

\bibitem{Dq1} A. D. Mirlin and F. Evers, Multifractality and critical fluctuations at the Anderson transition, Phys. Rev. B {\bf 62}, 7920 (2000).

\bibitem{Dq2} J. Martin, I. Garcia-Mata, O. Giraud and B. Georgeot, Multifractal wave functions of simple quantum maps. Phys. Rev. E, {\bf 82}, 046206 (2010).

\bibitem{Renyi-jpa} S. Gnutzmann and K. Zyczkowski, R\'{e}nyi-Wehrl entropies as measures of localization in phase space, J. Phys. A: Math. Gen. {\bf 34}, 10123 (2001).

\bibitem{WangQ-Dicke} Q. Wang and M. Robnik, Statistical properties of the localization measure of chaotic eigenstates in the Dicke model, Phys. Rev. E {\bf 102}, 032212 (2020).



\bibitem{Vidal05}Jos I. Latorre, Romn Ors, Enrique Rico, and Julien Vidal, Entanglement entropy in the Lipkin-Meshkov-Glick model,
Phys. Rev. A {\bf71}, 064101(2005)

\bibitem{Gisin} Florian Frwis, Roman Schmied, and Nicolas Gisin, Tighter quantum uncertainty relations following from a general probabilistic bound,
Phys. Rev. A {\bf 92}, 012102(2015)







\end{thebibliography}

\end{document}